\documentclass[12pt]{article}

\usepackage{amsmath,amssymb,amsthm,amsxtra,overpic,bbm,bm,epsfig,subfigure,multirow}
\usepackage{color}

\usepackage{pdflscape}

\usepackage{comment}

\begin{document}

\begin{center}
{\Large \bf Renormalisation Group Corrections to the Littlest Seesaw Model and Maximal Atmospheric Mixing }
\end{center}

\vspace{0.1cm}

\begin{center}
{\bf Stephen F. King$^{1}$} \footnote{E-mail: king@soton.ac.uk} \;,\quad {\bf Jue Zhang$^{2}$} \footnote{E-mail: juezhang87@pku.edu.cn} \;, \quad {\bf Shun Zhou$^{3,2}$} \footnote{E-mail: zhoush@ihep.ac.cn}

\vspace{0.1cm}

{$^{1}$School of Physics and Astronomy, University of Southampton,\\
SO17 1BJ Southampton, United Kingdom \\
$^{2}$Center for High Energy Physics, Peking University, Beijing 100871, China\\
$^{3}$Institute of High Energy Physics, Chinese Academy of Sciences, Beijing 100049, China}
\end{center}

\vspace{0.5cm}

\begin{abstract}
The Littlest Seesaw (LS) model involves two right-handed neutrinos and a very constrained Dirac neutrino mass matrix, involving one texture zero and two independent Dirac masses, leading to a highly predictive scheme in which all neutrino masses and the entire PMNS matrix is successfully predicted in terms of just two real parameters. We calculate the renormalisation group (RG) corrections to the LS predictions, with and without supersymmetry, including also the threshold effects induced by the decoupling of heavy Majorana neutrinos both analytically and numerically. We find that the predictions for neutrino mixing angles and mass ratios are rather stable under RG corrections. For example we find that the LS model with RG corrections predicts close to maximal atmospheric mixing,
$\theta_{23}=45^\circ \pm 1^\circ$, in most considered cases, in tension with the latest NOvA results. The techniques used here apply to other seesaw models with a strong normal mass hierarchy.
\end{abstract}

\newpage

\section{Introduction}

Although it has been well established by neutrino oscillation experiments that neutrinos are massive particles and lepton flavors are significantly mixed \cite{nobel}, the dynamical origin of neutrino mass generation and lepton flavor mixing is yet unknown~\cite{XZbook,King:2013eh}. Among a number of theoretical models for tiny neutrino masses, the simplest and most elegant one should be the canonical seesaw model~\cite{Minkowski:1977sc, Yanagida:1979ss, Gell-Mann:1979ss, Glashow:1979ss, Mohapatra:1979ia}, in which the standard model (SM) is extended with right-handed neutrino singlets $N^{}_{i{\rm R}}$ and the gauge-invariant Lagrangian relevant for neutrino masses and lepton flavor mixing reads
\begin{eqnarray}
-{\cal L}^{}_{\rm m} = \overline{\ell^{}_{\rm L}} Y^{}_l H E^{}_{\rm R} + \overline{\ell^{}_{\rm L}} Y^{}_\nu \tilde{H} N^{}_{\rm R} + \frac{1}{2} \overline{N^c_{\rm R}} M^{}_{\rm R} N^{}_{\rm R} + {\rm h.c.} \; ,
\end{eqnarray}
where $\ell^{}_{\rm L}$ and $\tilde{H} \equiv {\rm i}\sigma^{}_2 H^*$ stand respectively for the left-handed lepton and Higgs doublets, $E^{}_{\rm R}$ and $N^{}_{\rm R}$ are the right-handed charged-lepton and neutrino singlets, $Y^{}_l$ and $Y^{}_\nu$ are the charged-lepton and Dirac neutrino Yukawa coupling matrices, $M^{}_{\rm R}$ is the Majorana mass matrix of right-handed neutrino singlets. After the Higgs field acquires its vacuum expectation value (vev), i.e., $v \equiv \langle H \rangle \approx 174~{\rm GeV}$, and the gauge symmetry is spontaneously broken, the charged-lepton and Dirac neutrino mass matrices are given by $M^{}_l = Y^{}_l v$ and $M^{}_{\rm D} = Y^{}_\nu v$, respectively. Consequently, the effective neutrino mass matrix is $M^{}_\nu \approx  M^{}_{\rm D} M^{-1}_{\rm R} M^{\rm T}_{\rm D}$ and the lightness of active neutrinos ${\cal O}(M^{}_\nu) \sim 0.1~{\rm eV}$ can be ascribed to the heaviness of right-handed Majorana neutrinos ${\cal O}(M^{}_{\rm R}) \sim 10^{14}~{\rm GeV}$, given ${\cal O}(M^{}_{\rm D}) \sim 100~{\rm GeV}$ at the electroweak scale.

However, the general seesaw model involves a large number of free parameters mainly arising from the Dirac neutrino Yukawa coupling matrix $Y^{}_\nu$ in the flavor basis, where the charged-lepton and right-handed neutrino mass matrices $M^{}_l = \widehat{M}^{}_l \equiv {\rm Diag}\{m^{}_e, m^{}_\mu, m^{}_\tau\}$ and $M^{}_{\rm R} = \widehat{M}^{}_{\rm R} \equiv {\rm Diag}\{M^{}_1, M^{}_2, M^{}_3\}$ are diagonal. In order to reduce the number of free parameters in a successful seesaw model, one may consider the so-called minimal version of only two right-handed neutrinos, which was first proposed by one of us in Refs.~\cite{King:1999mb, King:2002nf},
focussing on the decoupling case of $M^{}_3 \gg M^{}_2 > M^{}_1$, and with one texture zero
in the Dirac neutrino mass matrix $M^{}_{\rm D}$.
Therefore, the lightest neutrino is massless, namely, $m^{}_1 = 0$ in the case of normal neutrino mass hierarchy (NH, i.e., $m^{}_1 < m^{}_2 < m^{}_3$) and $m^{}_3 = 0$ in the case of inverted neutrino mass hierarchy (IH, i.e., $m^{}_3 < m^{}_1 < m^{}_2$). A further simplification of the minimal seesaw model has been considered by Frampton, Glashow and Yanagida~\cite{Frampton:2002qc}, who assume two texture zeros in the Dirac neutrino mass matrix $M^{}_{\rm D}$ and demonstrate that both neutrino masses and the cosmological matter-antimatter asymmetry can be explained in this economical setup via the seesaw and leptogenesis mechanisms~\cite{Fukugita:1986hr}. The phenomenology of the minimal seesaw model was subsequently fully explored in the
literature~\cite{Guo:2003cc, Ibarra:2003up, Mei:2003gn, Guo:2006qa, Antusch:2011nz,
Harigaya:2012bw, Zhang:2015tea}. In particular, the NH case in the Frampton-Glashow-Yanagida model has been shown to be already excluded by the latest neutrino oscillation data~\cite{Harigaya:2012bw, Zhang:2015tea}.

More recently, the Littlest Seesaw (LS) model was put forward in Refs.~\cite{King:2013iva, Bjorkeroth:2014vha, King:2015dvf,BAU,King:2016yvg}, where two right-handed neutrino singlets $N^{\rm atm}_{\rm R}$ and $N^{\rm sol}_{\rm R}$
are introduced into the SM and a simple but viable structure of the Dirac neutrino Yukawa coupling matrix is conjectured as
\begin{eqnarray}
{\bf Case~A}:~Y^{}_\nu = \begin{pmatrix}
0 & b e^{{\rm i}\eta/2} \\
a & n b e^{{\rm i}\eta/2} \\
a & (n-2) b e^{{\rm i}\eta/2}
\end{pmatrix} \quad {\rm or} \quad {\bf Case~B}:~Y^{}_\nu = \begin{pmatrix}
0 & b e^{{\rm i}\eta/2} \\
a & (n-2) b e^{{\rm i}\eta/2} \\
a & n b e^{{\rm i}\eta/2}
\end{pmatrix}
\label{eq:Ynu0}
\end{eqnarray}
with $a, b, \eta$ being three real parameters and $n$ an integer. In the flavor basis where $M^{}_l = \widehat{M}^{}_l$ and $\widehat{M}^{}_{\rm R} = {\rm Diag}\{M^{}_{\rm atm}, M^{}_{\rm sol}\}$ are diagonal, neutrino masses and lepton flavor mixing parameters at the electroweak scale $\Lambda^{}_{\rm EW} \sim \mathcal{O}(100~{\rm GeV})$ can be derived by diagonalizing the effective neutrino mass matrix $M^{}_\nu = Y^{}_\nu \widehat{M}^{-1}_{\rm R} Y^{\rm T}_\nu v^2$. The low-energy phenomenology in the LS model case A has been studied in detail
both numerically~\cite{King:2013iva,Bjorkeroth:2014vha} and analytically~\cite{King:2015dvf}, where it has been found that the best fit to experimental data of neutrino oscillations is obtained for $n = 3$ for a particular choice of phase $\eta \approx 2\pi /3$,
while for case B the preferred choice is for $n = 3$ and $\eta \approx -2\pi /3$ \cite{King:2013iva,King:2016yvg}.
The prediction for the baryon number asymmetry in our Universe via leptogenesis within case A is also studied~\cite{BAU}, while a successful realization of the flavor structure of $Y^{}_\nu$ for case B in Eq.~(\ref{eq:Ynu0}) through an $S^{}_4 \times U(1)$ flavor symmetry is recently achieved in Ref.~\cite{King:2016yvg}, where the symmetry fixes $n=3$ and $\eta = \pm 2\pi/3$.

With the parameters $n=3$ and $\eta = \pm 2\pi/3$ fixed, there are only two remaining real free Yukawa parameters
in Eq.~(\ref{eq:Ynu0}), namely $a,b$, so the LS predictions then
depend on only two real free input combinations $m^{}_a=a^2v^2/M^{}_{\rm atm}$ and $m^{}_b=b^2v^2/M^{}_{\rm sol}$, in terms of which all neutrino masses and the PMNS matrix are determined. For instance, if $m_a$ and $m_b$ are chosen to fix $m_2$ and $m_3$, then the entire PMNS mixing matrix, including phases, is determined with no free parameters.
It turns out that the LS model predicts close to maximal atmospheric mixing at the high scale,
$\theta_{23}\approx 46^\circ $
for case A , or $\theta_{23}\approx 44^\circ $ for case B~\cite{King:2016yvg}, where both
predictions are challenged by the latest NOvA results in the $\nu_{\mu}$ disappearance channel~\cite{nova}
which indicates that $\theta_{23}=45^\circ$
is excluded at the 2.5 $\sigma$ CL, although T2K measurements in the same channel continue to prefer maximal mixing
\cite{Abe:2014ugx}.

In view of the great simplicity and high predictivity of the LS model, we are well motivated to consolidate its theoretical predictions by investigating the renormalization-group (RG) running of neutrino masses and lepton flavor mixing parameters, which is necessary to be taken into account as the seesaw scale is as high as $\Lambda^{}_{\rm SS} = 10^{10 - 15}~{\rm GeV}$, close to the scale of grand unified theories $\Lambda^{}_{\rm GUT} = 2\times 10^{16}~{\rm GeV}$. In particular, the threshold effects caused by the decoupling of two heavy right-handed neutrinos are examined in an analytical way. We demonstrate that the predictions for neutrino mixing angles and CP-violating phases are rather stable against the radiative corrections.
For example, we find that the LS model including RG corrections for both cases A and B, both with and without supersymmetry,
predicts maximal atmospheric mixing in the range $\theta_{23}=45^\circ \pm 1^\circ$.
Both numerical and analytical calculations are implemented to understand our observations.
The results are expected to be indicative of a large class of seesaw models with a strong mass hierarchy that
predict close to maximal atmospheric mixing, so we conclude that RG corrections are not generally sufficient to rescue such models
if maximal atmospheric mixing becomes excluded.

The remaining part of our paper is organized as follows. In Sec. 2, the general formalism for RG running of neutrino parameters and the treatment of seesaw threshold effects are briefly reviewed. After a brief review on the basic idea of the LS model in Sec. 3, the radiative corrections are calculated and discussed in Sec. 4. Finally, we summarize our main results in Sec. 5.

\section{Renormalisation Group Running }

As is well known~\cite{Weinberg:1979sa}, neutrino masses and flavor mixing parameters at the low-energy scale are governed by the dimension-five Weinberg operator $\kappa ( \overline{\ell^{}_{\rm L}} \cdot \tilde{H})(\tilde{H}^{\rm T} \cdot \ell^c_{\rm L})/2$, which can be derived by integrating out the heavy Majorana neutrinos. The effective neutrino coupling matrix $\kappa$ is related to the neutrino mass matrix as $M^{}_\nu = \kappa v^2$ in SM or $M^{}_\nu = \kappa v^2 \sin^2\beta$ in the minimal supersymmetric standard model (MSSM), where $\tan\beta$ denotes the ratio between the vev's of two Higgs doublets in MSSM. If the mass spectrum of heavy Majorana neutrinos is not strongly hierarchical, it is an excellent approximation that all of them are simultaneously integrated out at a common seesaw scale, namely, $\Lambda^{}_{\rm SS} = M^{}_1 < M^{}_2 < M^{}_3$. In the case of $M^{}_1 \ll M^{}_2 \ll M^{}_3$, however, we have to decouple the heavy Majorana neutrino one by one and carefully deal with the matching between the effective theory below $M^{}_i$ (for $i = 1, 2, 3$) and the other one above. The detailed discussions on the RG running of neutrino parameters and threshold effects in the canonical seesaw model can be found in Refs.~\cite{King:2000hk,Antusch:2005gp, Mei:2005qp, Ohlsson:2013xva}.

Above the seesaw thresholds, the one-loop RG equations of model parameters have been derived in
Refs.~\cite{King:2000hk,Antusch:2005gp,Mei:2005qp} and are collected as below
\begin{eqnarray} \label{eq:Yl_above_threshold}
\frac{{\rm d}Y^{}_l}{{\rm d}t} &=& \left(\alpha^{}_l + C^l_l H^{}_l + C^\nu_l H^{}_\nu \right)Y^{}_l \; , \\
\frac{{\rm d}Y^{}_\nu}{{\rm d}t} &=& \left(\alpha^{}_\nu + C^l_\nu H^{}_l + C^\nu_\nu H^{}_\nu \right)Y^{}_\nu \; , \\
\frac{{\rm d}M^{}_{\rm R}}{{\rm d}t} &=& C^{}_{\rm R} \left[ M^{}_{\rm R} \left(Y^\dagger_\nu Y^{}_\nu\right) + \left(Y^\dagger_\nu Y^{}_\nu\right)^{\rm T} M^{}_{\rm R}\right] \; ,
\label{eq:RGE3}
\end{eqnarray}
where $t \equiv \ln(\mu/\Lambda^{}_{\rm EW})/(16\pi^2)$ with $\mu$ being the renormalization scale, $H^{}_f \equiv Y^{}_f Y^\dagger_f$ for $f = l, \nu, u, d$, and the relevant coefficients $(C^l_l, C^\nu_l, C^l_\nu, C^\nu_\nu, C^{}_{\rm R}) = (3/2, -3/2, -3/2, 3/2, 1)$ in the SM while $(C^l_l, C^\nu_l, C^l_\nu, C^\nu_\nu, C^{}_{\rm R}) = (3, 1, 1, 3, 2)$ in the MSSM. As indicated by Eq.~(3), even if we start with a diagonal matrix $Y^{}_l$ at the initial energy scale, it may become non-diagonal because of the contribution from $H^{}_\nu$. In this case, one has to diagonalize both $Y^{}_l$ and the effective neutrino mass matrix $M^{}_\nu$ to obtain lepton flavor mixing matrix and extract mixing parameters. The flavor-independent coefficients $\alpha^{}_l$ and $\alpha^{}_\nu$ in Eqs. (3), (4) and (5) read
\begin{eqnarray}
\alpha^{}_l &\equiv& {\rm Tr}\left(3 H^{}_u + 3H^{}_d + H^{}_l + H^{}_\nu\right) - \left(\frac{9}{4}g^2_1 + \frac{9}{4}g^2_2\right) \; , \\
\alpha^{}_\nu &\equiv& {\rm Tr}\left(3 H^{}_u + 3H^{}_d + H^{}_l + H^{}_\nu\right) - \left(\frac{9}{20}g^2_1 + \frac{9}{4}g^2_2\right) \; ,
\end{eqnarray}
in the SM; and the corresponding results in the MSSM are
\begin{eqnarray}
\alpha^{}_l &\equiv& {\rm Tr}\left(3H^{}_d + H^{}_l \right) - \left(\frac{9}{5}g^2_1 + 3g^2_2\right) \; , \\
\alpha^{}_\nu &\equiv& {\rm Tr}\left(3 H^{}_u + H^{}_\nu\right) - \left(\frac{3}{5}g^2_1 + 3g^2_2\right) \; .
\end{eqnarray}
The one-loop RG equations of gauge couplings $g^{}_1$ and $g^{}_2$ in the SM and MSSM can be found in the literature and should be solved together with those in Eqs.~(3)--(5). For later convenience, one can also define the effective neutrino coupling matrix $\kappa \equiv Y^{}_\nu M^{-1}_{\rm R} Y^{\rm T}_\nu$ above the seesaw thresholds and its RG equation can be obtained by using Eqs.~(3)--(5). More explicitly, we have
\begin{eqnarray}
\frac{{\rm d}\kappa}{{\rm d}t} = 2\alpha^{}_\nu \kappa + \left(C^l_\kappa H^{}_l + C^\nu_\kappa H^{}_\nu\right)\kappa + \kappa \left(C^l_\kappa H^{}_l + C^\nu_\kappa H^{}_\nu\right)^{\rm T} \; ,
\label{eq:kappa3}
\end{eqnarray}
where $(C^l_\kappa, C^\nu_\kappa) = (-3/2, 1/2)$ in the SM while $(C^l_\kappa, C^\nu_\kappa) = (1, 1)$ in the MSSM.

Below the seesaw scale, i.e., $\mu < M^{}_1$, the model parameters for leptons contain only $Y^{}_l$ and $\kappa$. In the effective theory, the one-loop RG equations are~\cite{Antusch:2003kp,Antusch:2005gp,Mei:2005qp}
\begin{eqnarray} \label{eq:Yl_below_threshold}
\frac{{\rm d}Y^{}_l}{{\rm d}t} &=& \left(\widehat{\alpha}^{}_l + C^l_l H^{}_l \right) Y^{}_l \; , \\
\frac{{\rm d}\kappa}{{\rm d}t} &=& \widehat{\alpha}^{}_\kappa \kappa + C^l_\kappa \left(H^{}_l \kappa + \kappa H^{\rm T}_l \right) \; ,
\label{eq:kappa1}
\end{eqnarray}
where the flavor-independent coefficients are defined as
\begin{eqnarray}
\widehat{\alpha}^{}_l &\equiv& {\rm Tr}\left(3 H^{}_u + 3H^{}_d + H^{}_l\right) - \left(\frac{9}{4}g^2_1 + \frac{9}{4}g^2_2\right) \; , \\
\widehat{\alpha}^{}_\kappa &\equiv& {\rm Tr}\left(6 H^{}_u + 6H^{}_d + 2H^{}_l\right) - \left(3g^2_2 - \lambda\right) \; ,
\end{eqnarray}
with $\lambda$ being the quartic Higgs coupling in the SM; and
\begin{eqnarray}
\widehat{\alpha}^{}_l &\equiv& {\rm Tr}\left(3H^{}_d + H^{}_l\right) - \left(\frac{9}{5}g^2_1 + 3g^2_2\right) \; , \\
\widehat{\alpha}^{}_\kappa &\equiv& {\rm Tr}\left(6 H^{}_u\right) - \left(\frac{6}{5}g^2_1 + 6g^2_2\right) \; ,
\end{eqnarray}
in the MSSM. Note that $C^l_l$ and $C^l_\kappa$ in Eqs.~(11) and (12) are the same as those in Eq.~(3) and (10). It is worthwhile to note that if $Y^{}_l$ is taken to be diagonal at $\mu = M^{}_1$, it remains to be diagonal all the way down to the electroweak scale, as indicated by Eq.~(11). Hence, the lepton flavor mixing parameters are solely determined by the effective neutrino coupling matrix $\kappa$.

Finally, we have to deal with the RG running between any two seesaw thresholds and specify the matching conditions. Between the $i$-th and $(i-1)$-th thresholds (namely, for $ M^{}_{i-1} < \mu < M^{}_i$), the effective neutrino mass matrix is given by~\cite{Antusch:2005gp,Mei:2005qp,Antusch:2002rr,Lindner:2005pk}
\begin{eqnarray}
M^{(i)}_\nu = v^2 \left[ \kappa^{(i)} + Y^{(i)}_\nu {M^{(i)}_{\rm R}}^{-1} {Y^{(i)}_\nu}^{\rm T}\right] \; ,
\label{eq:matching}
\end{eqnarray}
in the SM, while $v^2$ should be replaced by $v^2\sin^2\beta$ in the MSSM. Here $\kappa^{(i)}$ arises from the decoupling of the right-handed Majorana neutrinos of masses equal to or heavier than $M^{}_i$, while the second term in the parentheses on the right-hand side of Eq.~(\ref{eq:matching}) is obtained by manually removing the parameters corresponding to decoupled heavy neutrinos. It should be emphasized that in the SM the RG running behaviors of those two terms are governed by two different sets of RG equations, resulting in the so-called ``threshold effects". We will discuss such effects in detail in Section 4.

Since the hierarchical mass spectrum $M^{}_1 \ll M^{}_2 \ll M^{}_3$ is
assumed in the LS model and the contribution from the heaviest Majorana neutrino $N^{}_3$ to neutrino masses is negligible, we simply ignore the decoupling of $N^{}_3$ and consider the RG running started from the initial energy scale $\mu^{}_0 = \Lambda^{}_{\rm GUT}$, where the Dirac neutrino Yukawa coupling matrix takes either form given in Eq.~(\ref{eq:Ynu0}). Then the RG running and threshold effects characterized by $M^{}_2$ and $M^{}_1$ are treated as described above.

\section{The Littlest Seesaw Model}

Before considering the running effects in the LS model~\cite{King:2015dvf}, we briefly recall its predictions for neutrino masses and flavor mixing when ignoring the RG running. First of all, given $Y^{}_\nu$ in {\rm case A} in Eq.~(\ref{eq:Ynu0}) and
assuming
$M^{}_{\rm R} = {\rm Diag}\{M^{}_{\rm atm}, M^{}_{\rm sol}\}$, one can immediately get the effective neutrino mass matrix via the seesaw formula
\begin{eqnarray}
M^{{\rm A}}_\nu = m_a^{} \begin{pmatrix}
0 & 0 & 0 \\
0 & 1 & 1 \\
0 & 1 & 1
\end{pmatrix} + m_b^{} e^{i \eta}_{} \begin{pmatrix}
1 & n & n-2 \\
n & n^2_{} & n(n-2) \\
n-2 & n(n-2) & (n-2)^2_{}
\end{pmatrix} \; ,
\end{eqnarray}
where $m_a^{} = a^2 v^2/M_{\rm atm}^{}$ and $m_b^{} = b^2 v^2/M_{\rm sol}^{}$. Since $M^{}_l = {\rm Diag}\{m^{}_e, m^{}_\mu, m^{}_\tau\}$ is diagonal, where $m^{}_\alpha$ for $\alpha = e, \mu, \tau$ are the charged-lepton masses, neutrino masses $m^{}_i$ (for $i = 1, 2, 3$) and the lepton flavor mixing matrix $U$ can be found by diagonalizing $M^{\rm A}_\nu$, namely, $U^\dagger M^{\rm A}_\nu U^* = {\rm Diag}\{0, m^{}_2, m^{}_3\}$. In practice, we first perform a basis transformation via $M^\prime_\nu = U^\dagger_{\rm TB} M^{\rm A}_\nu U^*_{\rm TB}$, where $U^{}_{\rm TB}$ stands for the tri-bimaximal mixing pattern~\cite{TBM1, TBM2, TBM3, He:2003rm}
\begin{eqnarray}
U^{}_{\rm TB} = \begin{pmatrix}
\displaystyle \frac{2}{\sqrt{6}} & \displaystyle \frac{1}{\sqrt{3}} & 0 \\
\displaystyle -\frac{1}{\sqrt{6}} & \displaystyle \frac{1}{\sqrt{3}} & \displaystyle \frac{1}{\sqrt{2}} \\
\displaystyle \frac{1}{\sqrt{6}} & \displaystyle -\frac{1}{\sqrt{3}} & \displaystyle \frac{1}{\sqrt{2}}
\end{pmatrix} \; .
\label{eq:TBM}
\end{eqnarray}
After this transformation, we have
\begin{eqnarray}
M^\prime_\nu = \begin{pmatrix}
0 & 0 & 0 \\
0 & 3m^{}_b e^{{\rm i}\eta} & \sqrt{6} m^{}_b e^{{\rm i}\eta} (n-1) \\
0 & \sqrt{6} m^{}_b e^{{\rm i}\eta} (n-1) & 2\left[m^{}_a + m^{}_b e^{{\rm i}\eta} (n-1)^2\right]
\end{pmatrix}
\equiv \begin{pmatrix}
0 & 0 & 0 \\
0 & x & y \\
0 & y & z
\end{pmatrix} \; ,
\label{eq:xyz}
\end{eqnarray}
which can be further diagonalized by a rotation $U^{}_{\rm b}(\theta)$ in the $2$-$3$ complex plane. The corresponding rotation angle $\theta$ is given by $\tan 2\theta = 2|xy^* + yz^*|/(|z|^2 - |x|^2)$, and thus the mixing matrix is $U = U^{}_{\rm TB} U^{}_{\rm b}(\theta)$. As the lightest neutrino is massless, i.e., $m^{}_1 = 0$, the lepton flavor mixing matrix $U$ can be parametrized in terms of three mixing angles $\theta^{}_{ij}$ for $ij = 12, 13, 23$, one Dirac-type CP-violating phase $\delta$ and one Majorana-type CP-violating phase $\sigma$, namely,
\begin{eqnarray}
U = \begin{pmatrix}
c^{}_{12}c^{}_{13} & c^{}_{13} s^{}_{12} &  s^{}_{13}e^{-{\rm i}\delta} \\
-c^{}_{23}s^{}_{12}-c^{}_{12}s^{}_{13}s^{}_{23}e^{{\rm i}\delta} & c^{}_{12}c^{}_{23}-s^{}_{12}s^{}_{13}s^{}_{23}e^{{\rm i}\delta} & c^{}_{13}s^{}_{23} \\
s^{}_{12}s^{}_{23}-c^{}_{12}c^{}_{23}s^{}_{13}e^{{\rm i}\delta} & -c^{}_{12}s^{}_{23}-c^{}_{23}s^{}_{12}s^{}_{13}e^{{\rm i}\delta} & c^{}_{13}c^{}_{23}
\end{pmatrix}
\begin{pmatrix}
1 & 0 & 0 \\
0 & e^{{\rm i} \sigma} & 0 \\
0 & 0 & 1
\end{pmatrix} \; ,
\end{eqnarray}
where $c^{}_{ij} \equiv \cos \theta^{}_{ij}$ and $s^{}_{ij} \equiv \sin \theta^{}_{ij}$ have been defined. As shown in Ref.~\cite{King:2015dvf}, neutrino masses $\{m^{}_2, m^{}_3\}$, flavor mixing angles $\{\theta^{}_{12}, \theta^{}_{13}, \theta^{}_{23}\}$, and CP-violating phases $\{\delta, \sigma\}$ can be exactly calculated in terms of the model parameters $m^{}_a$, $m^{}_b$ and $\eta$.

However, in the sequential-dominance approximation, implying $m^{}_a \gg m^{}_b$ and $|z| \gg |x|, |y|$, the neutrino masses turn out to be
\begin{eqnarray}
m^{}_1 = 0 \;, \quad m^{}_2 \approx 3 m^{}_b \; , \quad m^{}_3 \approx 2 m^{}_a \; ,
\end{eqnarray}
while the mixing angles are
\begin{eqnarray}
\sin \theta^{}_{13} &\approx& \frac{\tan 2\theta}{2\sqrt{3}} \; , \quad \tan \theta^{}_{12} = \frac{1}{\sqrt{2}} \left(1 - 3\sin^2 \theta^{}_{13}\right)^{1/2} \; , \quad \tan \theta^{}_{23} \approx 1 + \frac{2\tan 2\theta}{\sqrt{6}} \cos \omega \; ,\nonumber \\
\end{eqnarray}
where $\tan 2\theta \approx \sqrt{6} m^{}_b(n-1)/\left|m^{}_a + m^{}_b e^{{\rm i}\eta} (n-1)^2\right|$ and $\omega = \arg\left[m^{}_a + m^{}_b e^{{\rm i}\eta} (n-1)^2\right] - \eta$. It is worthwhile to notice that the correlation between $\theta^{}_{12}$ and $\theta^{}_{13}$ in the above equation is exact, as a salient feature of the LS model. In addition, two CP-violating phases are~\cite{King:2015dvf}
\begin{eqnarray}
\sin \delta &\approx& - \frac{24m^3_a m^3_b (n-1)}{m^2_2 m^2_3 \Delta m^2_{32} s^{}_{12} c^{}_{12} s^{}_{23} c^{}_{23} s^{}_{13} c^2_{13}} \sin \eta\; , \nonumber \\
\sin \sigma &\approx& +\frac{m^{}_a m^{}_b \left[4m^2_a - m^2_b (2n+1)^2 (n-2)^2\right]}{m^{}_2 m^{}_3 \Delta m^2_{32} c^2_{12} c^2_{13} c^2_{23} s^2_{23}} \sin \eta\; ,
\end{eqnarray}
where $\Delta m^2_{ji} \equiv m^2_j - m^2_i$ is the neutrino mass-squared difference.

If the form of $Y^{}_\nu$ in case B in Eq.~(\ref{eq:Ynu0}) is taken, the corresponding effective neutrino mass matrix $M^{\rm B}_\nu$ is related to that in case A via $M^{\rm B}_\nu = P^{}_{23} M^{\rm A}_\nu P^{\rm T}_{23}$, where $P^{}_{23}$ denotes the elementary transformation matrix that exchanges the second and third columns or rows of an arbitrary $3\times 3$ matrix. While $M^{\rm A}_\nu = U\cdot{\rm Diag}\{0, m^{}_2, m^{}_3\} \cdot U^{\rm T}$ has been archived, we immediately arrive at $M^{\rm B}_\nu = (P^{}_{23} U) \cdot{\rm Diag}\{0, m^{}_2, m^{}_3\} \cdot (P^{}_{23}U)^{\rm T}$. Then, it is straightforward to verify that such a transformation leads to the following relations between two sets of mixing parameters
\begin{eqnarray}
\theta^{\rm B}_{12} = \theta^{\rm A}_{12} \; , \quad \theta^{\rm B}_{13} = \theta^{\rm A}_{13} \; , \quad \theta^{\rm B}_{23} = \frac{\pi}{2} - \theta^{\rm A}_{23} \; , \quad \delta^{\rm B} = \pi -\delta^{\rm A} \; , \quad \sigma^{\rm B} = \pi - \sigma^{\rm A}  \; .
\end{eqnarray}
Therefore, it is unnecessary to explicitly diagonalize $M^{\rm B}_\nu$, and all the mixing parameters can be calculated by using the above relations while neutrino mass eigenvalues remain the same. Some comments on the model predictions are in order:
\begin{itemize}
\item Two predictive ans\"{a}tze of $Y^{}_\nu$ with $n = 3$ will be considered. The first one is $Y^{}_\nu$ in case A, and $\eta = 2\pi/3$, together with $m^{}_a = 25.67~{\rm meV}$ and $m^{}_b = 2.684~{\rm meV}$, is assumed. One can exactly diagonalize $M^{\rm A}_\nu$ and find out neutrino masses $\{m^{}_1, m^{}_2, m^{}_3\} = \{0, 8.59, 49.8\}~{\rm meV}$, $\{\theta^{}_{12}, \theta^{}_{13}, \theta^{}_{23}\} = \{34.3^\circ, 8.67^\circ, 45.8^\circ\}$ and $\delta = -86.7^\circ$, which are in perfect agreement with the global-fit results~\cite{global1, global2, global3} for $m^{}_1 = 0$. The second one is $Y^{}_\nu$ in case B with $n = 3$ and $\eta = -2\pi/3$, and the same parameters $m^{}_a = 25.67~{\rm meV}$ and $m^{}_b = 2.684~{\rm meV}$ are adopted. Consequently, we have $M^{\rm B}_\nu = P^{}_{23} {M^{\rm A}_\nu}^* P^{\rm T}_{23}$, implying the same mass eigenvalues, $\{\theta^{}_{12}, \theta^{}_{13}, \theta^{}_{23}\} = \{34.3^\circ, 8.67^\circ, 44.2^\circ\}$ and $\delta = -93.3^\circ$~\cite{King:2016yvg}. The prediction for $\delta = -86.7^\circ$ or $-93.3^\circ$ is compatible with the recent hints from T2K and NOvA experiments on a nearly maximal CP-violating phase.

\item As $m^{}_1 = 0$ is implied in the LS, neutrino mass hierarchy is obviously normal. In this case, the effective neutrino mass for neutrinoless double-beta decays is as small as $m^{}_{\beta \beta} = m^{}_b = 2.684~{\rm meV}$, which is impossible to measure in the foreseeable future. These conclusions are applicable to both case A with $\eta = 2\pi/3$ and case B with $\eta = -2\pi/3$.

\item For the chosen input parameters, the baryon number asymmetry is found to be $Y^{}_{\rm B} \approx 8.4 \times 10^{-11}$ can be reproduced for $M^{}_1 \approx 3.9\times 10^{10}~{\rm GeV}$~\cite{Bjorkeroth:2014vha}. As indicated by Eq.~(24), the CP violation in neutrino oscillations and that for the cosmological matter-antimatter asymmetry are determined by the same parameter $\eta$. For a different value of $\eta$, the heavy neutrino masses $M^{}_1$ and $M^{}_2$ can be changed by choosing suitable parameters $a$ and $b$, without spoiling the low-energy predictions for neutrino masses and mixing angles.
\end{itemize}
Since the seesaw scale is extremely high, one may be worried about whether the RG running effects can significantly modify the above conclusions. This problem will be addressed in the following section.

\section{Renormalisation Group Corrections to the Littlest Seesaw Model Predictions}

The RG running effects on neutrino mixing parameters in the SM and in the NH case are expected to be rather small. However, the strongly hierarchical mass spectrum $M^{}_1 \ll M^{}_2$ implies that the seesaw threshold effects can be important, depending on the flavor structure of Dirac neutrino Yukawa coupling matrix $Y^{}_\nu$. On the other hand, if the LS model is supersymmetrized, a large value of $\tan\beta$ leads to an increase of charged-lepton Yukawa couplings, which may enhance the RG running effects. Therefore, we are motivated to carry out a detailed study of those effects. Since the analysis is almost identical for both case A and B,
for definiteness we only consider the RG corrections in full detail for one of the two cases,
namely case A, then later highlight the differences which are important for case B.

\subsection{Case A from $\Lambda^{}_{\rm GUT}$ to $M^{}_2$}

First of all, we need to specify the input parameters at the initial scale $\mu^{}_0 = \Lambda^{}_{\rm GUT}$. In this subsection, we focus on the form of $Y^{}_\nu$ in case A, and the other scenario will be considered later.
We first consider the mass ordering of right-handed neutrinos
$M_1= M_{\rm atm}$, $M_2= M_{\rm sol}$, where by definition $M_1< M_2$.
Later we shall consider the results for the alternative mass ordering.
Note that the low energy effective neutrino mass matrix is independent of this heavy right-handed neutrino mass
ordering, but the RG corrections in the heavy threshold region dependent on it.

To be consistent with the consequential dominance, we take $M^{}_1 = 10^{12}~{\rm GeV}$ and $M^{}_2 = 10^{15}~{\rm GeV}$ for illustration, implying $M^{}_2 \gg M^{}_1$. Furthermore, as shown in the previous section, the global-fit results of neutrino mixing parameters can be well reproduced for $n = 3$, together with $m^{}_a = 25.67~{\rm meV}$, $m^{}_b = 2.684~{\rm meV}$ and $\eta = 2\pi/3$. In this case, $Y^{}_\nu(\mu^{}_0)$ is given by Eq.~(\ref{eq:Ynu0}) with $a \approx 0.03$ and $b \approx 0.3$, satisfying $b \gg a$. Therefore, it is interesting to notice a strong hierarchy among the matrix elements of $Y^{}_\nu$, and $|\left(Y^{}_\nu\right)^{}_{\mu 2}| = 3b$ is the largest one. Note that such a choice of $Y_\nu^{}(\mu_0^{})$ also implies that we are in the flavor basis at this initial boundary scale, namely, both $Y_l^{}(\mu_0^{})$ and $M_{\mathrm{R}}^{}(\mu_0^{})$ are diagonal.

Since the lepton flavor mixing matrix arises from the mismatch between the diagonalization of charged-lepton Yukawa matrix $Y_l^{}$ and that of the neutrino mass matrix $M_\nu^{}$, we therefore pay particular attention to the RG running of both $Y_l^{}$ and $M_\nu^{}$. It is well known that below the seesaw threshold (i.e., $\mu < M_1^{}$), $Y_l^{}$ would always stay diagonal at one-loop level if it is diagonal initially at the boundary [see Eq.~(\ref{eq:Yl_below_threshold})]. However, this is no longer the case when considering the RG running above the seesaw threshold, due to the term involving $Y_\nu^{}$ in Eq.~(\ref{eq:Yl_above_threshold}). In the following we then trace the evolution of both $Y_l^{}$ and $M_\nu^{}$ analytically for the running between $\Lambda_{\mathrm{GUT}}^{}$ and $M_2^{}$.

Let us start with the RG running of $M_\nu^{}$.  Above the seesaw threshold $\mu = M^{}_2$, the RG running of the would-be neutrino mass matrix $M^{}_\nu \equiv \kappa v^2$ in the SM (or $M^{}_\nu \equiv \kappa v^2 \sin^2\beta$ in the MSSM) is governed by Eq.~(\ref{eq:kappa3}). Neglecting the relatively small contribution from $Y_l^{}$, the evolution of the flavor structure in $M_\nu^{}$  is mainly driven by the term involving
\begin{eqnarray}
H^{}_\nu \equiv Y^{}_\nu Y^\dagger_\nu \approx \begin{pmatrix}
0 & 0 & 0\\
0 & 9 b^2_{} & 0\\
0 & 0 & 0
\end{pmatrix} \; ,
\label{eq:Hnu}
\end{eqnarray}
where the approximation $a \ll b \ll 3b$ has been made in $Y^{}_\nu$ to simplify our analytical discussions. If the second column of $Y^{}_\nu$ is fully kept, $H^{}_\nu$ will be a $3\times 3$ real and symmetric matrix without any vanishing elements, and it is difficult to deal with the radiative corrections to neutrino mixing angles in an analytical way. In the approximation made in Eq.~(\ref{eq:Hnu}), it is straightforward to solve the Eq.~(\ref{eq:kappa3}) and obtain
\begin{eqnarray}
M_\nu^{} (t) = I_\alpha^{} \begin{pmatrix}
1 & 0 & 0\\
0 & I_\nu^{} & 0\\
0 & 0 & 1
\end{pmatrix} M_\nu^0 \begin{pmatrix}
1 & 0 & 0\\
0 & I_\nu^{} & 0 \\
0 & 0 & 1
\end{pmatrix},
\end{eqnarray}
where all the parameters at the initial scale $\mu^{}_0 = \Lambda^{}_{\rm GUT}$ are denoted by a subscript or superscript $``0"$. More explicitly, we have defined $M_\nu^{0} \equiv M_\nu^{}(t_0^{})$, and the evolution functions $I_\alpha^{}$ and $I_\nu^{}$ are found to be
\begin{eqnarray}
I_\alpha^{} &=& \text{exp} \left[\int_{t_0^{}}^t 2\alpha_\nu^{} (t^\prime)~{\rm d} t^\prime \right] \; , \\
I_\nu^{} &=& \text{exp} \left[\int_{t_0^{}}^t \frac{9}{2}b(t^\prime)^2_{} ~{\rm d} t^\prime  \right] \; .
\end{eqnarray}
Assuming that $b(t)$ does not run much from the initial value $b^{}_0 = b(t^{}_0)$ = 0.3, then we have $I_\nu^{} \approx 1 - \epsilon^{}_\nu$ with $\epsilon^{}_\nu \equiv 9 b^2_0 (t_0^{} - t)/2$. At the threshold $\mu = M^{}_2 = 10^{15}~{\rm GeV}$, one can obtain $t^{}_0 - t = \ln(\Lambda^{}_{\rm GUT}/M^{}_2)/(16\pi^2) \approx 0.02$ and thus $\epsilon^{}_\nu \approx 7.7\times 10^{-3}$, which serves as an excellent perturbation parameter. Therefore, we arrive at
\begin{eqnarray} \label{eq:pertb_1}
M_\nu^{}(t)/I_\alpha^{} = M_\nu^0 - \epsilon^{}_\nu
\begin{pmatrix}
0 & (M_\nu^0)_{e\mu} & 0 \\
(M_\nu^0)_{\mu e} & 2 (M_\nu^0)_{\mu \mu} & (M_\nu^0)_{\mu \tau} \\
0 & (M_\nu^0)_{\tau \mu} & 0
\end{pmatrix} + \mathcal{O}(\epsilon^2_\nu) \; ,
\label{eq:mnu}
\end{eqnarray}
with $(M^0_\nu)^{}_{\alpha \beta}$ for $\alpha, \beta = e, \mu, \tau$ being the matrix elements of $M^0_\nu$. It is interesting to note that the one-loop RG corrections to $M_\nu^{}$ are quite similar to those for $\kappa$ below the seesaw threshold, where the dominant corrections from the tau Yukawa coupling $y_\tau^{}$ modify the third row and column of $\kappa$.

To extract the RG corrections to three mixing angles, we have to diagonalize the mass matrix in Eq.~(\ref{eq:mnu}). This can be achieved perturbatively in two steps. First, as shown in the previous section, the leading-order mass matrix $M^0_\nu$ can be diagonalized by a unitary matrix $U^{\nu}_0 = U^{}_{\rm TB} U^{}_{\rm b}$, namely, $U^{\nu\dagger}_0 M^0_\nu U^{\nu*}_0 = D^{\nu}_0 \equiv \mathrm{Diag}\{0, m^0_2, m^0_3\}$ with both $m_2^0$ and $m_3^0$ real and positive. Here $U_{\rm TB}^{}$ is the tri-bimaximal mixing matrix given in Eq.~(\ref{eq:TBM}), and
\begin{eqnarray}
U_{\rm b}^{} = \begin{pmatrix}
1 & & \\
& e^{{\rm i} \varphi}_{} & \\
& & 1
\end{pmatrix}
\begin{pmatrix}
1 & 0 & 0 \\
0 & \cos\theta  & \sin\theta  \\
0 & -\sin\theta  & \cos\theta
\end{pmatrix}
\begin{pmatrix}
1 & & \\
& e^{{\rm i} \phi_2^{}/2}_{} & \\
& & e^{ {\rm i} \phi_3^{}/2}_{}
\end{pmatrix} \; .
\end{eqnarray}
Both $\varphi$ and $\theta$ can be obtained by diagonalizing $M_{\rm b} M_{\rm b}^\dagger$ with $M_{\rm b} \equiv U_{\rm TB}^\dagger M_\nu^0 U_{\rm TB}^*$, namely, $\varphi = \arg[xy^* + yz^*]$ and $\tan 2\theta = 2|xy^* + yz^*|/(|z|^2 - |x|^2)$, where $x, y, z$ have been introduced in Eq.~(\ref{eq:xyz}). In addition, $\phi_2^{}$ and $\phi_3^{}$ are obtained by requiring both $m_2^0$ and $m_3^0$ to be real and positive. More details on the diagonalization of $M^0_\nu$ can be found in the previous section and in Ref.~\cite{King:2015dvf}.

Second, after the unitary transformation $U^{\nu\dagger}_0 [M^{}_\nu(t)I^{-1}_\alpha] U^{\nu*}_0 = M^{\nu}_{\rm p}$, we are left with a mass matrix $M^{\nu}_{\rm p}$, which is almost diagonal except for small corrections proportional to $\epsilon$ in both diagonal and off-diagonal entries. As we are interested in the radiative corrections to neutrino mixing angles, it is sufficient to find out a unitary matrix that diagonalize $M^{\nu}_{\rm p} M^{\nu\dagger}_{\rm p}$, i.e., $U^{\nu\dagger}_{\rm p} (M^{\nu}_{\rm p} M^{\nu\dagger}_{\rm p}) U^{\nu*}_{\rm p} = {\rm Diag}\{0, m^2_2, m^2_3\}$. The unitary matrix $U^{\nu}_{\rm p}$ can be found by using the standard perturbation theory~\cite{Sakurai}, and the final mixing matrix in the neutrino sector is given by $U_\nu^{} = U^{\nu}_0 U^{\nu}_{\rm p}$ up to one physical Majorana-type CP-violating phase.

Having obtained the mixing matrix $U_\nu^{}$ for the neutrino mass matrix $M_\nu^{}$ at $M_2^{}$, we then focus on the mixing matrix from the charged-lepton Yukawa matrix $Y_l^{}$. To this end, we first study the evolution of $Y_l^{}$ from $\Lambda_{\mathrm{GUT}}^{}$ to $M_2^{}$ with the help of Eq.~(\ref{eq:Yl_above_threshold}). Unlike the above discussions on the evolution of $M_\nu^{}$, now we can keep all the non-zero elements in $H_\nu^{}$, owing to the simpler structure of the RG equation for $Y_l^{}$, namely,
\begin{eqnarray}
\frac{{\rm d} Y_l^{}}{{\rm d} t} = \left[ \alpha_l^{} -\frac{3}{2} b^2_{}
\begin{pmatrix}
1 & 3 & 1 \\
3 & 9 & 3 \\
1 & 3 & 1
\end{pmatrix} \right ] Y_l^{} \; .
\end{eqnarray}
Since we are interested in the flavor mixing induced by $Y_l^{}$, the first term with a flavor-independent coefficient $\alpha_l^{}$ in the above RG equation can be neglected. Then, we solve it analytically, with a diagonal form of $Y_l^{}$ at the high-energy boundary $\Lambda_{\mathrm{GUT}}^{}$. In view of the strong hierarchy $y^2_e \ll y^2_\mu \ll y^2_\tau$, the unitary matrix $U^{}_l$ defined by $U_l^\dagger Y_l^{} Y_l^\dagger U_l^{} = \mathrm{Diag}\{y_e^2, y_\mu^2, y_\tau^2\}$ turns out to be
\begin{eqnarray}
U_l^{} \approx {\bf 1} + \epsilon^{}_l \begin{pmatrix}
0 & 3  & \sqrt{10} \sin\theta_l^{} \\
-3 & 0 & \sqrt{10} \cos\theta_l^{} \\
-\sqrt{10} \sin\theta_l^{} & -\sqrt{10} \cos\theta_l^{} & 0
\end{pmatrix} \; ,
\end{eqnarray}
where $\epsilon_l^{} \equiv 3 b^2_0 (t^{}_0 - t)/2$ serves as another small parameter for expansion, and $\theta_l^{} = \arctan(3/4)/2$ stems from the diagonalization of $H_\nu^{}$, which is needed to solve Eq.~(32) analytically.

With $U_l^{}$ in Eq.~(33), we obtain the lepton mixing matrix $U_{}^{} = U_l^\dagger U_\nu^{}$ at the scale of $M_2^{}$. Thus, three neutrino mixing angles at $\mu = M^{}_2$ can be exacted in the leading-order approximation
\begin{eqnarray}
\theta_{13}^\prime &\approx & \theta_{13}^0 - \epsilon^{}_\nu \left[ \frac{ m^0_2}{\sqrt{3} m^0_3}\cos(\varphi + \phi_2^{} - \phi_3^{}) - \frac{\sqrt{3}}{4} \theta_{13}^0 \cot \theta^0_{12} \right] \cos \theta^0_{12} \; \nonumber \\
& ~& ~~~~~~~~~~~~~~~~ \quad - ~ \epsilon_l^{} \left[ 3 \cos(\delta^0 + \rho) \sin\theta_{23}^0  + \sqrt{10} \cos\theta_{23}^0 \cos(\delta^0 - \rho) \sin\theta_l^{}\right] \; , \nonumber \\
\tan \theta_{12}^\prime & \approx & \frac{1}{\sqrt{2}} - \frac{3(\theta_{13}^{0})^2}{2\sqrt{2}} + \frac{\epsilon^{}_\nu}{2\sqrt{2}} - \frac{3\epsilon^{}_l}{2} \left[3 \cos\theta_{23}^0 \cos\rho - \sqrt{10} \cos\rho \sin\theta_{23}^0 \sin\theta_l^{} \right]  \; ,  \label{eq:approx_1} \\
\tan \theta_{23}^\prime & \approx & \tan \theta_{23}^0 - \frac{\sqrt{3}}{2\sqrt{2}} \sec^2\theta_{23}^{0} \cos \theta_{12}^0  ~\epsilon^{}_\nu  -\sqrt{10} \cos(2\rho) \cos\theta_l^{} \sec^2\theta_{23}^0 ~ \epsilon_l^{} \nonumber \; ,
\end{eqnarray}
where $\{\theta_{12}^\prime, \theta_{13}^\prime, \theta_{23}^\prime\}$ stand for the mixing angles at $\mu = M^{}_2$, while $\{\theta_{12}^0, \theta_{13}^0, \theta_{23}^0, \delta_{}^0\}$ for those at $\mu^{}_0 = \Lambda^{}_{\rm GUT}$. In addition, the approximations $\tan \theta^0_{12} \approx [1 - 3 (\theta^0_{13})^2/2]/\sqrt{2}$ and $\sin \theta \approx \sqrt{3} \sin \theta^0_{13}/(\sqrt{2} \tan \theta^0_{12})$ have been made, and $\rho \equiv \arg\left[ \cos \theta/\sqrt{2} + \sin \theta e^{{\rm i} \varphi}/\sqrt{3}\right]$ has been defined.

Numerically, we have made a comparison between the results from the analytical formulas and those from solving exactly the RG equations. For illustration, the RG running effects in the SM case are considered. The final results are shown in Table~\ref{tb:validation}. As one can see, the approximate formulas in Eq.~(\ref{eq:approx_1}) yield very good predictions (namely, the fifth row in Table~\ref{tb:validation}) for $\theta^{}_{13}$ and $\theta^{}_{23}$. However, for $\theta^{}_{12}$, we obtain a slightly larger value, which can be ascribed to the rough approximation at the very beginning, namely, keeping only the dominant term in $H_\nu^{}$. To see this point clearly, we calculate the mixing angles directly from Eq.~(\ref{eq:pertb_1}), while the exact $U_l^{}$ is obtained from the actual RG running, and show the numerical results in the fourth row of Table~\ref{tb:validation}. An excellent agreement between the values in the fourth and fifth rows validates the above perturbation method leading to Eq.~(\ref{eq:approx_1}).

\begin{table}
\centering
\begin{tabular}{c | c | c | c}
\hline
\hline
 & $\theta_{13}^{}$(deg) & $\theta_{12}^{}$(deg) & $\theta_{23}^{}$(deg) \\
\hline
Exact, at $\mu_0^{}$ & 8.67 & 34.32 & 45.77 \\
Exact, at $M_2^{}$ & 8.57 & 34.04 & 44.89 \\
Exact, at $M_1^{}$ & 8.11 & 34.13 & 44.39 \\
\hline
Approximation in Eq.~(\ref{eq:pertb_1}), at $M_2^{}$ & 8.58 & 34.18 & 44.79 \\
Approximation in Eq.~(\ref{eq:approx_1}), at $M_2^{}$ & 8.59 & 34.18 & 44.80 \\
\hline
Approximation in Eq.~(\ref{eq:approx_2}), at $M_1^{}$ & 8.01 & 34.11 & 44.49 \\
\hline
\hline
\end{tabular}
\caption{Comparison between the numerical and analytical results of three neutrino mixing angles at different energy scales. The initial values at $\mu^{}_0 = \Lambda^{}_{\rm GUT} = 2\times 10^{16}~{\rm GeV}$ are given in the first row, while the exact values at $\mu = M^{}_2 = 10^{15}~{\rm GeV}$ and $\mu = M^{}_1 = 10^{12}~{\rm GeV}$ are calculated by solving the full set of RG equations and are listed in the second and third rows, respectively. The approximate analytical results are shown in the last three rows.}
\label{tb:validation}
\end{table}

\subsection{Case A from $M^{}_2$ to $M^{}_1$}

In this subsection we proceed with case A to consider
the threshold effects due to the decoupling of heavy right-handed neutrinos on the neutrino mixing angles. Since $M^{}_2$ is very close to $\Lambda^{}_{\rm GUT}$, it is reasonable to assume that the RG running effects of both $Y^{}_\nu$ and $M^{}_{\rm R}$ in the first stage are negligible. Therefore, we have the following light neutrino mass matrix at $\mu = M_2^{}$, which can be decomposed into two terms
\begin{eqnarray} \label{eq:mnuM2}
M_\nu^\prime =  \widetilde{M}_\nu^{} + v^2 \widehat{\kappa} \; ,
\end{eqnarray}
with $\widetilde{M}_\nu^{} \equiv v^2 \widetilde{Y}_\nu^{} {M}_{1}^{-1} \widetilde{Y}_\nu^{\rm T}$ and $\widehat{\kappa} \equiv \widehat{Y}_\nu M_2^{-1} \widehat{Y}_\nu^{\rm T}$, where $\widetilde{Y}_{\nu}$ and $\widehat{Y}_\nu$ stand for the first and second columns of $Y_\nu^{}$ given in Eq.~(\ref{eq:Ynu0}), respectively. As we have shown in Sec. 2, in the SM the RG equations of $\widetilde{M}_\nu^{}$ and $\widehat{\kappa}$ in the effective theory after the decoupling of $N^{}_2$ have different coefficients from the Higgs self-coupling and gauge coupling contributions, leading to significant threshold effects for a hierarchical mass spectrum of heavy neutrinos. In the MSSM, one has to replace $v^2$ with $v^2 \sin^2 \beta$ in Eq.~(32), but both $\widetilde{M}_\nu^{}$ and $\widehat{\kappa}$ evolve in the same way, which is not very interesting in view of threshold effects (although in the RG equations we need to use $\widetilde{Y}_\nu^{}$ instead of $Y_\nu^{}$). For this reason, we focus on the case of SM.

For clarity, we recap the RG equations of $\widetilde{M}_\nu^{}$ and $\widehat{\kappa}$ in the SM, which have already been given in Eqs.~(\ref{eq:kappa3}) and (\ref{eq:kappa1}) and can be expressed as follows
\begin{eqnarray}
\frac{{\rm d} X}{{\rm d} t} = \widetilde{\alpha}_X^{}  X + \left (-\frac{3}{2} H_l^{} + \frac{1}{2} \widetilde{H}_\nu^{} \right ) X + X \left (-\frac{3}{2} H_l^{} + \frac{1}{2} \widetilde{H}_\nu^{} \right)^{\rm T}_{},
\end{eqnarray}
for $X = \widehat{\kappa}$ or $\widetilde{M}_\nu^{}$. Here $\widetilde{H}_\nu^{} = \widetilde{Y}_\nu \widetilde{Y}^\dagger_\nu$, and $\widetilde{\alpha}_X^{}$ is given by
\begin{eqnarray}
\widetilde{\alpha}_\kappa &=& 2\mathrm{Tr}[3H_u^{} + 3H_d^{} + H_l^{} + \widetilde{H}_\nu^{}] + \lambda - 3 g_2^2 \; , \\
\widetilde{\alpha}_\nu &=& 2\mathrm{Tr}[3H_u^{} + 3H_d^{} + H_l^{} + \widetilde{H}_\nu^{}] - \frac{9}{10} g_1^2 - \frac{9}{2} g_2^2 \; ,
\end{eqnarray}
In the case of $M_1^{} = 10^{12}~\text{GeV}$ and $M_2^{} = 10^{15}~\text{GeV}$ under discussion, all three entries in $\widetilde{Y}_\nu^{}$ are quite small, we thus neglect both $H_l^{}$ and $\widetilde{H}_\nu^{}$ in the RG equations for both $\widetilde{M}_\nu^{}$ and $\widehat{\kappa}$. As an immediate consequence, the running of $\widetilde{M}_\nu$ and $\widehat{\kappa}$ only differ in the flavor-independent coefficient $\widetilde{\alpha}_X^{}$. Following Ref.~\cite{Bergstrom:2010id}, we can obtain the neutrino mass matrix $M_\nu^{\prime\prime}$ at $\mu = M_1^{}$ as
\begin{eqnarray}
M_\nu^{\prime\prime} \approx \zeta (M_\nu^\prime + \xi v^2_{} \widehat{\kappa}) \; ,
\end{eqnarray}
with
\begin{eqnarray}
\zeta \approx \left( \frac{M_1^{}}{M_2^{}} \right)^{\widetilde{\alpha}_\nu^{}/16\pi^2} \; , \quad
\xi \approx \left(\frac{M_1^{}}{M_2^{}} \right)^{(\widetilde{\alpha}^{}_\kappa -\widetilde{\alpha}_\nu^{})/16\pi^2} - 1 \; .
\end{eqnarray}
Hence, the radiative corrections to three neutrino mixing angles come from the $\xi v^2 \widehat{\kappa}$ term, which reflects how large the running effects are between $M_2^{}$ and $M_1^{}$.

To start with, we can diagonalize the neutrino mass matrix at $M_2^{}$ via a unitary transformation, i.e., $M_\nu^\prime = U^\prime_{} D_\nu^\prime {U^\prime}^{\rm T}_{}$ with $D_\nu^\prime = \mathrm{Diag}\{0, m_2^\prime, m_3^\prime\}$. The unitary matrix $U^\prime$ is given by
\begin{eqnarray}
U^\prime_{} =
P^{}_\rho
\begin{pmatrix}
c_{12}^\prime c_{13}^\prime & c_{13}^\prime s_{12}^\prime &  s_{13}^\prime e^{-i\delta^\prime} \\
-c_{23}^\prime s_{12}^\prime -c_{12}^\prime s_{13}^\prime s_{23}^\prime e^{i\delta^\prime} & c_{12}^\prime c_{23}^\prime -s_{12}^\prime s_{13}^\prime s_{23}^\prime e^{i\delta^\prime} & c_{13}^\prime s_{23}^\prime \\
s_{12}^\prime s_{23}^\prime -c_{12}^\prime c_{23}^\prime s_{13}^\prime e^{i\delta^\prime} & -c_{12}^\prime s_{23}^\prime -c_{23}^\prime s_{12}^\prime s_{13}^\prime e^{i\delta^\prime} & c_{13}^\prime c_{23}^\prime
\end{pmatrix}
P^{}_\omega \; ,
\end{eqnarray}
where $P^{}_\rho \equiv {\rm Diag}\{e^{{\rm i} \rho_1^\prime} , e^{{\rm i} \rho_2^\prime}, e^{{\rm i} \rho_3^\prime}\}$ and $P^{}_\omega \equiv {\rm Diag}\{1, e^{{\rm i} \omega_2^\prime}, 1\}$ are diagonal phase matrices, $s_{ij}^\prime \equiv \sin\theta_{ij}^\prime$ and $c_{ij}^\prime \equiv \cos\theta_{ij}^\prime$ for $ij = 12, 13, 23$ have been defined. Here the symbols with a single prime indicate the parameters at the scale of $M_2^{}$, while those with double primes are the parameters at $M_1^{}$. Note that there in general will be two Majorana-type CP-violating phases in the last matrix on the right-hand side of Eq.~(38), of which however only one is physical because of one massless neutrino (i.e., $m^{}_1 = 0$).

Next, we consider the flavor structure of $\widehat{\kappa}$, which is reconstructed by the second column of $Y^{}_\nu$ and $M^{}_2$. Keeping all the elements of $\widehat{\kappa}$, one can obtain
\begin{eqnarray} \label{eq:pertb_2}
\frac{M_\nu^{\prime\prime}}{\zeta m_3^\prime} =  \frac{ M_\nu^\prime }{m_3^\prime} + \widehat{\epsilon} e^{{\rm i} \eta} \begin{pmatrix}
1 & 3 & 1 \\
3 & 9 & 3 \\
1 & 3 & 1
\end{pmatrix} \; ,
\end{eqnarray}
where $\widehat{\epsilon} = b^2 v^2 \xi/(M_2^{} m_3^\prime)$ would be a small expansion parameter. We then diagonalize $M_\nu^{\prime \prime}$ perturbatively following the same procedure as that in the previous section. In the final step, we multiply the obtained mixing matrix $U_{\nu}^{\prime\prime}$ from the $M_\nu^{\prime\prime}$ by the previously found $U_l^{}$ so as to construct the flavor mixing matrix at $M_1^{}$, i.e., $U(M_1^{}) = U_l^\dagger U_\nu^{\prime\prime}$. Note that $Y_l^{}$ does not run much from $M_2^{}$ to $M_1^{}$. The three mixing angles are then extracted as
\begin{eqnarray} \label{eq:approx_2}
\theta_{13}^{\prime\prime} &\approx & \theta_{13}^\prime - \left[ 3 \cos(\delta^\prime - \gamma^\prime_{12}) s_{23}^\prime + \sqrt{10} \cos(\delta^\prime - \gamma^\prime_{13}) c_{23}^\prime \sin\theta_l^{} \right] \epsilon_l^{} \nonumber \\
&~& + \left[ \cos\beta_{13}^\prime c_{23}^\prime + 3\cos\beta_{12}^\prime s_{23}^\prime \right] \widehat{\epsilon}  \; , \nonumber \\
t_{12}^{\prime\prime} &\approx& t_{12}^\prime - \left[ 3 c_{23}^\prime \cos\gamma^\prime_{12} - \sqrt{10} \cos\gamma^\prime_{13} s_{23}^\prime \sin\theta_l^{} \right] \frac{\epsilon_l^{}}{c^{\prime 2}_{12}}  + \frac{m^\prime_3}{m^\prime_2} [ 3 c_{23}^\prime \cos\alpha_{12}^\prime - \cos\alpha_{13}^{\prime} s_{23}^\prime ] c_{12}^\prime \widehat{\epsilon} \nonumber \\
&~& + \frac{m^\prime_3}{m^\prime_2} \left[ \cos\alpha_{11}^\prime - 9 c_{23}^{\prime 2} \cos\alpha_{22}^{\prime 2} - s_{23}^{\prime 2} \cos\alpha_{33}^\prime  - s_{12}^{\prime 2} t_{12}^\prime (3 c_{23}^\prime \cos\alpha_{12}^\prime - \cos\alpha_{13}^{\prime} s_{23}^\prime) \right. \nonumber \\
&~& \left. + 6s^\prime_{23} c^\prime_{23} \cos\alpha_{23}^\prime \right] t_{12}^\prime \widehat{\epsilon} - \left\{ 3 c_{23}^{\prime 2} \cos\beta_{23}^\prime + c_{23}^\prime (9 \cos\beta_{22}^\prime - \cos\beta_{33}^\prime ) s_{23}^\prime \right. \nonumber \\
&~& \left. - 3 s_{23}^\prime \left[ \cos\beta_{23}^\prime  s_{23}^\prime + ( \cos\beta_{13}^\prime - \cos\beta_{12}^\prime ) c^{\prime 2}_{12} t_{12}^\prime \right] \right\} \frac{\theta_{13}^\prime \widehat{\epsilon}}{c^{\prime 2}_{12}} \; , \\
t_{23}^{\prime\prime} &\approx & t_{23}^\prime - \sqrt{10} \cos\gamma^\prime_{23} \frac{\cos\theta_l^{}}{c^{\prime 2}_{23}} ~\epsilon_l^{} +\left\{ 3 (1- t_{23}^{\prime 2}) \cos(\alpha^\prime_{23} + 2\omega^\prime_2) \right. \nonumber \\
&~& \left. + \left[9 \cos(\eta- 2\rho_2^\prime) - \cos(\eta - 2\rho_3^\prime) \right] s_{23}^\prime c_{23}^\prime \right\} \widehat{\epsilon} \; , \nonumber
\end{eqnarray}
where $t^\prime_{ij} \equiv \tan \theta^\prime_{ij}$ and $t^{\prime \prime}_{ij} \equiv \tan \theta^{\prime \prime}_{ij}$ have been introduced for $ij = 12, 23$, and $\alpha_{ij}^\prime \equiv \eta -\rho_i^\prime - \rho_j^\prime - 2\omega_2^\prime$, $\beta_{ij}^\prime \equiv \delta^\prime  + \eta -\rho_i^\prime - \rho_j^\prime$, and $\gamma^\prime_{ij} \equiv \rho^\prime_i - \rho^\prime_j$ have been defined for $i, j = 1, 2, 3$. Note that only the leading-order contributions from $\widehat{\epsilon}$, $m^\prime_2/m^\prime_3$ and $\theta_{13}^\prime$ are kept in Eq.~(\ref{eq:approx_2}), except that for $\theta_{12}^{\prime\prime}$ we also include corrections of the order of $\theta_{13}^\prime \widehat{\epsilon}$ for better accuracy.

Numerical verification of our approximate formulas is also given in Table~\ref{tb:validation}. Using the exact results of three mixing angles at $M_2^{}$ as input, we compute the approximate results at $M_1^{}$ from Eq.~(\ref{eq:approx_2}), which have been shown in the last row. In comparison with the exact results in the third row, we can observe that the approximate formulas indeed capture the major threshold effects.

\begin{figure}[!t]
\begin{center}
\includegraphics[width=\textwidth]{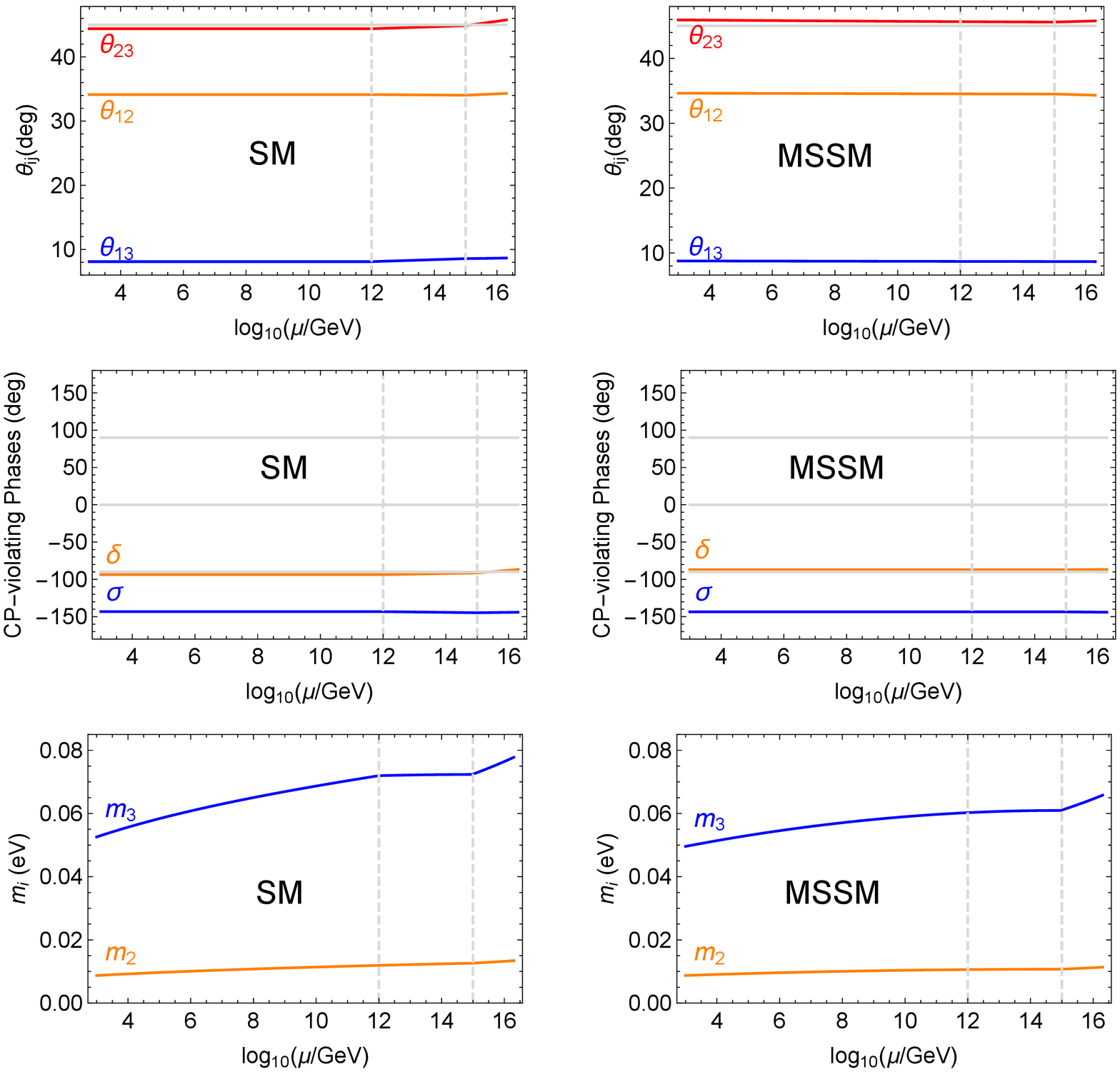}
\end{center}
\vspace{-0.3cm}
\caption{The evolution of three mixing angles $\{\theta^{}_{12}, \theta^{}_{13}, \theta^{}_{23}\}$, two CP-violating phases $\{\delta, \sigma\}$ and neutrino masses $\{m^{}_2, m^{}_3\}$ within the SM (left) and MSSM with $\tan\beta = 30$ (right) for the form of $Y_\nu^{}$ in {\bf Case A} given in Eq.~(\ref{eq:Ynu0}). The initial values for the most relevant parameters at the high-energy scale $\mu^{}_0$ include $g_1^{} = 0.579$, $g_2^{} = 0.521$, $g_3^{} = 0.527$, $\lambda = 0.5$ (only for the SM), $y^{}_\tau = 0.010$ and $y_t^{} = 0.483$.}
\label{fg:fig1}
\end{figure}

As is well known, the running effects of neutrino mixing parameters below the seesaw scale $\mu = M^{}_1$ are insignificant, in particular for the NH case. On the other hand, even in the leading-order approximation, it is complicated to derive any analytical results for the CP-violating phases and neutrino masses. Therefore, in order to fully address the RG running effects from $\Lambda^{}_{\rm GUT} = 2\times 10^{16}~{\rm GeV}$ to $\Lambda^{}_{\rm EW} = 10^3~{\rm GeV}$, we numerically solve the full set of RG equations with the $\texttt{REAP}$ package \cite{Antusch:2005gp} for three neutrino mixing angles $\{\theta^{}_{12}, \theta^{}_{13}, \theta^{}_{23}\}$, two CP-violating phases $\{\delta, \sigma\}$, and two neutrino masses $\{m^{}_2, m^{}_3\}$.  The final results are depicted in Fig.~\ref{fg:fig1} (together with numerical values at various energy scales in Table~\ref{tb:tb2}), and the main features are summarized as follows: (1) All the mixing angles and CP-violating phases are rather stable against the RG corrections. The largest deviation from the initial value is observed for $\theta^{}_{23}$, but even in this case the deviation is only around one degree. Therefore, the theoretical predictions for mixing angles and CP-violating phases in the LS model can be applied at both low- and high-energy scales. (2) However, it should be noticed that the running of absolute neutrino masses is remarkable. To be consistent with neutrino oscillation data, the initial values of $Y^{}_\nu$ should be multiplied by a factor of $1.25$ (or $1.15$) for SM (or MSSM), which has already been taken into account in Fig.~\ref{fg:fig1}. This overall scaling of $Y^{}_\nu$ does not alter the results for three flavour mixing angles at the high-energy boundary, but it does modify the absolute values of $Y^{}_\nu$, leading to slightly larger $a$ and $b$.

\begin{table}
\footnotesize
\centering
\begin{tabular}{c | c  c  c  c | c  c  c  c | c}
\hline
\hline
& \multicolumn{4}{| c |}{SM} & \multicolumn{4}{ c|}{MSSM ($\tan\beta = 30$)} & \multirow{2}{*}{Best fit} \\
\cline{1-9}
& $\Lambda_{\mathrm{GUT}}^{}$ & $M_2^{}$ & $M_1^{}$ & $\Lambda_{\mathrm{EW}}^{}$ & $\Lambda_{\mathrm{GUT}}^{}$ & $M_2^{}$ & $M_1^{}$ & $\Lambda_{\mathrm{EW}}^{}$ & \\
\hline
$\theta_{13}^{} (\mathrm{deg})$ & 8.67 & 8.57 & 8.11 & 8.11 & 8.67 & 8.68 & 8.70 & 8.77 & $8.46^{+0.14}_{-0.15}$\\
$\theta_{12}^{} (\mathrm{deg})$ & 34.32 & 34.04 & 34.13 & 34.13 & 34.32 & 34.50 & 34.53 &  34.63 & $33.72^{+0.79}_{-0.76}$\\
$\theta_{23}^{} (\mathrm{deg})$ & 45.77 & 44.89 & 44.40 & 44.40 & 45.77  & 45.60 & 45.66 & 45.92  & $41.5^{+1.3}_{-1.1}$\\
\hline
$\delta (\mathrm{deg})$ & $- 86.7$ & $-91.4$ & $-93.7$ & $-93.7$ & $-86.7$ & $-87.0$ & $-87.0$ &  $-87.0$ & $-71^{+38}_{-51}$\\
$\sigma (\mathrm{deg})$ & $-144.0$ & $-144.7$ & $-143.2$ & $-143.2$ & $-144.0$ & $-143.5$ & $-143.5$ &  $-143.5$ & --\\
\hline
$m_2^{} (\mathrm{meV})$ & 13.4 & 12.6 & 11.9 & 8.72 & 11.4 & 10.7 & 10.6 &  8.74 & $8.65^{+0.11}_{-0.09}$\\
$m_3^{} (\mathrm{meV})$ & 77.8 & 72.4 & 72.0 & 52.6 & 65.8 & 61.1 & 60.3 &  49.6 & $50.26^{+0.39}_{-0.37}$\\
$m_2^{}/m_3^{}$ & 0.172 & 0.174 & 0.165 & 0.166 & 0.173 & 0.175 & 0.176 & 0.176 & $0.172^{+0.003}_{-0.003}$ \\
\hline
\hline
\end{tabular}
\caption{Three mixing angles $\{\theta^{}_{12}, \theta^{}_{13}, \theta^{}_{23}\}$, two CP-violating phases $\{\delta, \sigma\}$ and non-zero light neutrino masses $\{m^{}_2, m^{}_3\}$ at various energy scales according to two scenarios in {\bf Case A} given in Fig.~\ref{fg:fig1}. For comparison, we also show the best-fit results from Ref.~\cite{global1} in the last column.}
\label{tb:tb2}
\end{table}

\subsection{Case B}

We now discuss case B in Eq.~(\ref{eq:Ynu0}) with $n = 3$ and $\eta = -2\pi/3$. It has been found \cite{King:2016yvg, King:2013iva} that this alternative scenario of $Y^{}_\nu$ also yields a phenomenologically successful and predictive description of neutrino masses and lepton mixing parameters, if RG corrections are ignored~\cite{King:2016yvg}. Following a similar treatment as in the previous case, we now study the RG running effects given this new form of $Y_\nu^{}$. The analytical formulas for flavour mixing angles are almost the same as before, except for two differences.

(1) During the running from $\Lambda_{\mathrm{GUT}}^{}$ to $M_2^{}$, we shall take a form of $H_\nu^{}$ as
\begin{eqnarray}
H_\nu^{} \approx \begin{pmatrix}
0 & 0 & 0 \\
0 & 0 & 0 \\
0 & 0 & 9 b^2_{}
\end{pmatrix} \; ,
\end{eqnarray}
instead of that in Eq.~(\ref{eq:Hnu}). Consequently, in order to obtain $M_\nu^{}$ at $M_2^{}$, we need to consider corrections to the third row and column of $M_\nu^0$ at $\Lambda_{\mathrm{GUT}}^{}$. In this case, we have to replace Eq.(~\ref{eq:mnu}) with the following
\begin{eqnarray}
M_\nu^{}(t)/I_\alpha^{} = M_\nu^0 - \epsilon^{}_\nu
\begin{pmatrix}
0 & 0 & (M_\nu^0)_{e\tau} \\
0 & 0 & (M_\nu^0)_{\mu \tau} \\
(M_\nu^0)_{\tau e} & (M_\nu^0)_{\tau \mu} & 2 (M_\nu^0)_{\tau \tau}
\end{pmatrix} + \mathcal{O}(\epsilon^2_\nu) \; .
\end{eqnarray}
Adopting the previous diagonalization procedure, we find that the analytical formulas for $\theta_{13}^\prime$ and $\theta_{12}^\prime$ remain the same as those in Eq.~(\ref{eq:approx_1}), while for $\theta_{23}^\prime$ we have
\begin{eqnarray}
\tan \theta_{23}^\prime & \approx & \tan \theta_{23}^0 + \frac{\sqrt{3}}{2\sqrt{2}} \sec^2\theta_{23}^{0} \cos \theta_{12}^0  ~\epsilon^{}_\nu  -\sqrt{10} \cos 2\rho \cos\theta_l^{} \sec^2\theta_{23}^0 ~ \epsilon_l^{} \; ,
\end{eqnarray}
where all the parameters follow the same definitions as in the previous subsections. It is worthwhile to point out that the correction proportional to $\epsilon^{}_\nu$ in the above equations has an opposite sign to that in Eq.~(\ref{eq:approx_1}), which can be used to explain the difference between the running behavior of decreasing $\theta^{}_{23}$ in case A and that of increasing $\theta^{}_{23}$ in case B.

(2) For threshold effects arising from the running between $M_2^{}$ and $M_1^{}$, the modification on the previous analytical study shows up in Eq.~(\ref{eq:pertb_2}), namely,
\begin{eqnarray}
\frac{M_\nu^{\prime\prime}}{\zeta m_3^\prime} =  \frac{ M_\nu^\prime }{m_3^\prime} + \widehat{\epsilon} e^{{\rm i} \eta} \begin{pmatrix}
1 & 1 & 3 \\
1 & 1 & 3 \\
3 & 3 & 9
\end{pmatrix} \; .
\end{eqnarray}
It is straightforward to verify that such a modification leads to slightly different analytical formulas for three flavour mixing angles:
\begin{eqnarray} \label{eq:approx_3}
\theta_{13}^{\prime\prime} &\approx & \theta_{13}^\prime - \left[ 3 \cos(\delta^\prime - \gamma^\prime_{12}) s_{23}^\prime + \sqrt{10} \cos(\delta^\prime - \gamma^\prime_{13}) c_{23}^\prime \sin\theta_l^{} \right] \epsilon_l^{} \nonumber \\
&~& + \left( 3\cos\beta_{13}^\prime c_{23}^\prime + \cos\beta_{12}^\prime s_{23}^\prime \right) \widehat{\epsilon}  \; , \nonumber \\
t_{12}^{\prime\prime} &\approx& t_{12}^\prime - \left[ 3 c_{23}^\prime \cos\gamma^\prime_{12} - \sqrt{10} \cos\gamma^\prime_{13} s_{23}^\prime \sin\theta_l^{} \right] \frac{\epsilon_l^{}}{c^{\prime 2}_{12}}  + \frac{m^\prime_3}{m^\prime_2} [ c_{23}^\prime \cos\alpha_{12}^\prime - 3\cos\alpha_{13}^{\prime} s_{23}^\prime ] c_{12}^\prime \widehat{\epsilon} \nonumber \\
&~&  + \frac{m^\prime_3}{m^\prime_2} \left[ 6s^\prime_{23} c^\prime_{23} \cos\alpha_{23}^\prime  -  c_{23}^{\prime 2} \cos\alpha_{22}^{\prime 2} - 9s_{23}^{\prime 2} \cos\alpha_{33}^\prime - s_{12}^{\prime 2} t_{12}^\prime ( c_{23}^\prime \cos\alpha_{12}^\prime - 3\cos\alpha_{13}^{} s_{23}^\prime) \right. \nonumber \\
&~& \left. + \cos\alpha_{11}^\prime \right] t_{12}^\prime \widehat{\epsilon} - \left[ 3 c_{23}^{\prime 2} \cos\beta_{23}^\prime + c_{23}^\prime (\cos\beta_{22}^\prime - 9\cos\beta_{33}^\prime ) s_{23}^\prime - 3 s_{23}^\prime \cos\beta_{23}^\prime  s_{23}^\prime \right ] \frac{\theta_{13}^\prime \widehat{\epsilon}}{c^{\prime 2}_{12}} \; , \nonumber \\
t_{23}^{\prime\prime} &\approx & t_{23}^\prime - \sqrt{10} \cos\gamma^\prime_{23} \frac{\cos\theta_l^{}}{c^{\prime 2}_{23}} ~\epsilon_l^{} + \left\{ 3 (1- t_{23}^{\prime 2}) \cos(\alpha^\prime_{23} + 2\omega^\prime_2) \right. \nonumber \\
&~& \left. + \left[\cos(\eta- 2\rho_2^\prime) - 9\cos(\eta - 2\rho_3^\prime)\right] s_{23}^\prime c_{23}^\prime \right\} \widehat{\epsilon} \; ,
\end{eqnarray}
where the relevant parameters have been defined below Eq.~(\ref{eq:approx_2}). Comparing between Eq.~(\ref{eq:approx_2}) and Eq.~(\ref{eq:approx_3}), one can observe that only the coefficients in front of a few terms are different.

\begin{figure}[!t]
\begin{center}%
\includegraphics[width=\textwidth]{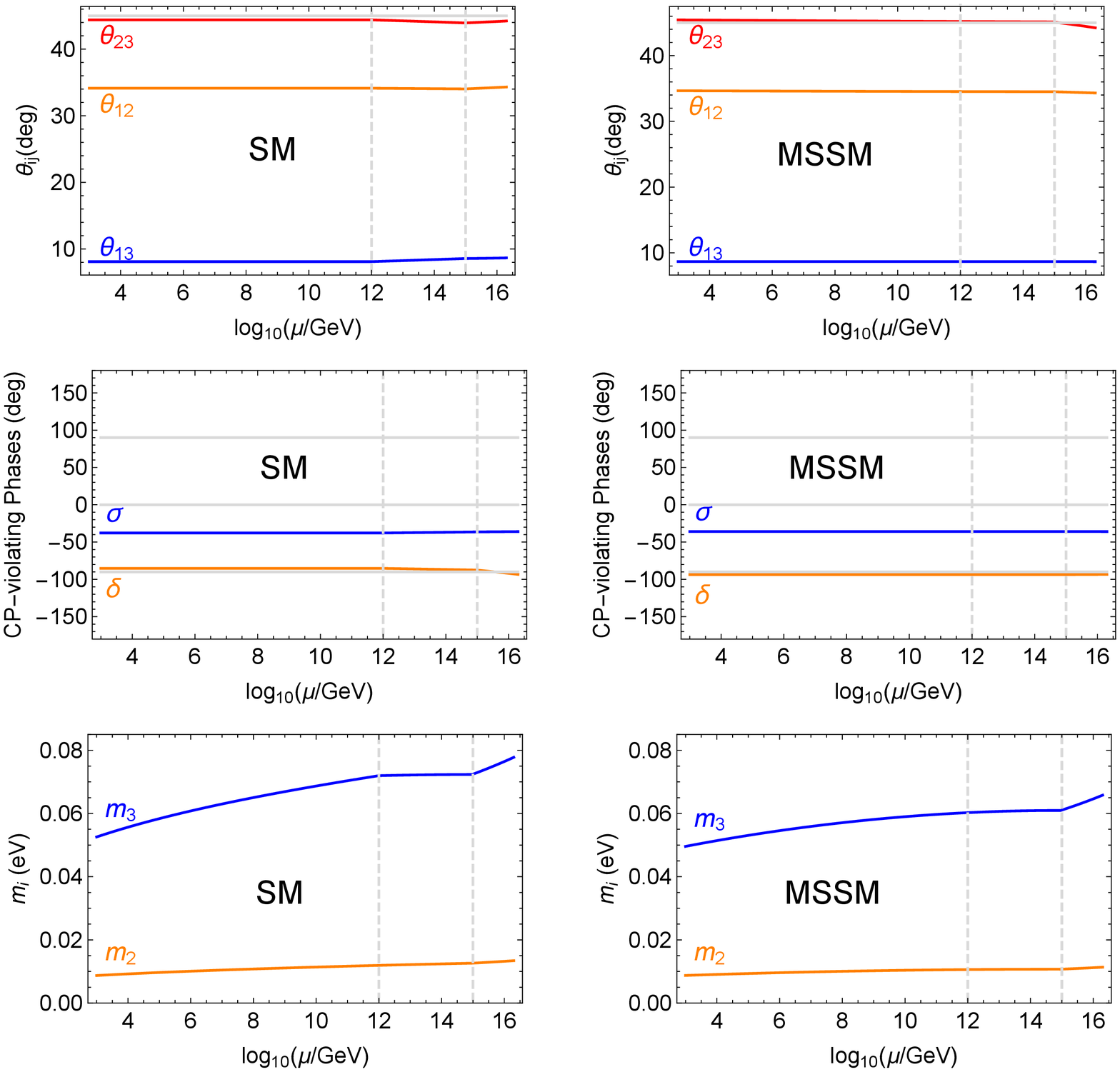}
\end{center}
\vspace{-0.3cm}
\caption{The evolution of three mixing angles $\{\theta^{}_{12}, \theta^{}_{13}, \theta^{}_{23}\}$, two CP-violating phases $\{\delta, \sigma\}$ and neutrino masses $\{m^{}_2, m^{}_3\}$ within the SM (left) and MSSM with $\tan\beta = 30$ (right) for the form of $Y_\nu^{}$ in {\bf Case B} given in Eq.~(\ref{eq:Ynu0}). The initial values for the most relevant parameters at the high-energy scale $\mu^{}_0$ include $g_1^{} = 0.579$, $g_2^{} = 0.521$, $g_3^{} = 0.527$, $\lambda = 0.5$ (only for the SM), $y^{}_\tau = 0.010$ and $y_t^{} = 0.483$.}
\label{fg:fig2}
\end{figure}

Numerical RG evolution of this alternative form of $Y_\nu^{}$ is also performed in Fig.~\ref{fg:fig2}, with the same input parameters as those in Fig.~\ref{fg:fig1} except for the sign of $\eta$. Also, we show the detailed numerical values for three mixing angles, two CP-violating phases and neutrino masses at various energy scales in Table \ref{tb:tb3}. As one can see, RG corrections to mixing angles and phases are quite stable as in the previous case, and similar running behaviours are also observed for neutrino masses.

\begin{table}[!t]
\small
\centering
\begin{tabular}{c | c  c  c  c | c  c  c  c | c}
\hline
\hline
& \multicolumn{4}{| c |}{SM} & \multicolumn{4}{ c|}{MSSM ($\tan\beta = 30$)} & \multirow{2}{*}{Best fit} \\
\cline{1-9}
& $\Lambda_{\mathrm{GUT}}^{}$ & $M_2^{}$ & $M_1^{}$ & $\Lambda_{\mathrm{EW}}^{}$ & $\Lambda_{\mathrm{GUT}}^{}$ & $M_2^{}$ & $M_1^{}$ & $\Lambda_{\mathrm{EW}}^{}$ & \\
\hline
$\theta_{13}^{} (\mathrm{deg})$ & 8.67 & 8.57 & 8.11 & 8.11 & 8.67 & 8.67 & 8.67 & 8.67 & $8.46^{+0.14}_{-0.15}$\\
$\theta_{12}^{} (\mathrm{deg})$ & 34.32 & 34.03 & 34.13 & 34.13 & 34.32 & 34.50 & 34.54 &  34.65 & $33.72^{+0.79}_{-0.76}$\\
$\theta_{23}^{} (\mathrm{deg})$ & 44.22 & 43.94 & 44.40 & 44.40 & 44.22  & 45.10 & 45.19 & 45.43  & $41.5^{+1.3}_{-1.1}$\\
\hline
$\delta (\mathrm{deg})$ & $-93.3$ & $-87.6$ & $-85.2$ & $-85.2$ & $-93.3$ & $-93.6$ &  $-93.6$ & $-93.7$ & $-71^{+38}_{-51}$\\
$\sigma (\mathrm{deg})$ & $-36.0$ & $-36.4$ & $-37.8$ & $-37.8$ & $-36.0$ & $-35.9$ & $-35.9$ &  $-35.9$ & --\\
\hline
$m_2^{} (\mathrm{meV})$ & 13.4 & 12.6 & 11.9 & 8.72 & 11.4 & 10.7 & 10.6 &  8.74 & $8.65^{+0.11}_{-0.09}$\\
$m_3^{} (\mathrm{meV})$ & 77.8 & 72.4 & 72.0 & 52.6 & 65.8 & 61.1 & 60.3 &  49.6 & $50.26^{+0.39}_{-0.37}$\\
$m_2^{}/m_3^{}$ & 0.172 & 0.174 & 0.165 & 0.166 & 0.173 & 0.175 & 0.176 & 0.176 & $0.172^{+0.003}_{-0.003}$ \\
\hline
\hline
\end{tabular}
\caption{Three mixing angles $\{\theta^{}_{12}, \theta^{}_{13}, \theta^{}_{23}\}$, two CP-violating phases $\{\delta, \sigma\}$ and non-zero light neutrino masses $\{m^{}_2, m^{}_3\}$ at various energy scales according to two scenarios in {\bf Case B} given in Fig.~\ref{fg:fig2}. For comparison, we also show the best-fit results from Ref.~\cite{global1} in the last column.}
\label{tb:tb3}
\end{table}

\subsection{Alternative Ordering of $M_{\rm atm}^{}$ and $M_{\rm sol}^{}$}

In the previous discussions, we have assumed the mass matrix of heavy right-handed neutrinos to be $M^{}_{\rm R} = {\rm Diag}\{M^{}_{\rm atm}, M^{}_{\rm sol}\}$ and taken the normal mass ordering as $M_{\rm atm}^{} = M^{}_1 = 10^{12}~{\rm GeV}$ and $M_{\rm sol}^{} = M_2^{} = 10^{15}~{\rm GeV}$. As we have mentioned, there exists an alternative ordering, namely, $M_{\rm atm}^{} = M_2^{} = 10^{15}~{\rm GeV}$ and $M_{\rm sol}^{} = M_1^{} = 10^{12}~{\rm GeV}$. In this case, in order to obtain the same neutrino masses and mixing angles as before, we require $m^{}_a = a^2 v^2/M^{}_{\rm atm} = 25.67~{\rm meV}$ and $m^{}_b = b^2 v^2/M^{}_{\rm sol} = 2.684~{\rm meV}$, implying $a \approx 0.94$ and $b \approx 0.01$. Although neutrino masses and mixing angles are kept unchanged, the RG running and threshold effects should be quite different for the following reasons:
\begin{enumerate}
\item Now that the mass ordering of two heavy Majorana neutrinos is inverted, we have to exchange the two columns of $Y^{}_\nu$ in Eq.~(\ref{eq:Ynu0}), namely,
\begin{eqnarray}
{\bf Case~C}: Y^{}_\nu = \begin{pmatrix}
b e^{{\rm i}\eta/2} & 0 \\
n b e^{{\rm i}\eta/2} & a \\
(n-2) b e^{{\rm i}\eta/2} & a
\end{pmatrix} ~~~~~~ {\rm or} ~~~~~~ {\bf Case~D}: Y^{}_\nu = \begin{pmatrix}
b e^{{\rm i}\eta/2} & 0 \\
(n-2) b e^{{\rm i}\eta/2} & a \\
n b e^{{\rm i}\eta/2} & a
\end{pmatrix}
\; .
\label{eq:Ynu0I}
\end{eqnarray}
When crossing the seesaw thresholds, we first decouple the heaviest neutrino at $M^{}_2$ (by ignoring the second column of $Y^{}_\nu$ for $\mu < M^{}_2$), and then the second one at $M^{}_1$. It is evident that the flavour structure of $Y^{}_\nu$ at each stage is distinct from that for the normal ordering.

\item During the running from $\Lambda^{}_{\rm GUT}$ to $M^{}_2$, the evolution of neutrino mixing angles is mainly governed by
    \begin{eqnarray}
    H^{}_\nu \approx \begin{pmatrix}
                0 & 0 & 0\\
                0 & a^2 & a^2 \\
                0 & a^2 & a^2
                \end{pmatrix}
                \; ,
    \label{eq:HnuI}
    \end{eqnarray}
    where the dominant element $a^2 \approx 0.88$ is much larger than the others. Moreover, $H^{}_\nu$ is not diagonal, and thus affects greatly the flavour structure of $M^{}_\nu$. For the same reason, it seems impossible to solve the RG equation of $M^{}_\nu$ analytically.

\item During the running from $M^{}_2$ to $M^{}_1$, the reduced Yukawa coupling matrix involves only the parameter $b \approx 0.01$, which is much smaller than that in the previous case. Therefore, we expect insignificant running effects from the neutrino sector.
\end{enumerate}

\begin{figure}[!t]
\begin{center}
\includegraphics[width=\textwidth]{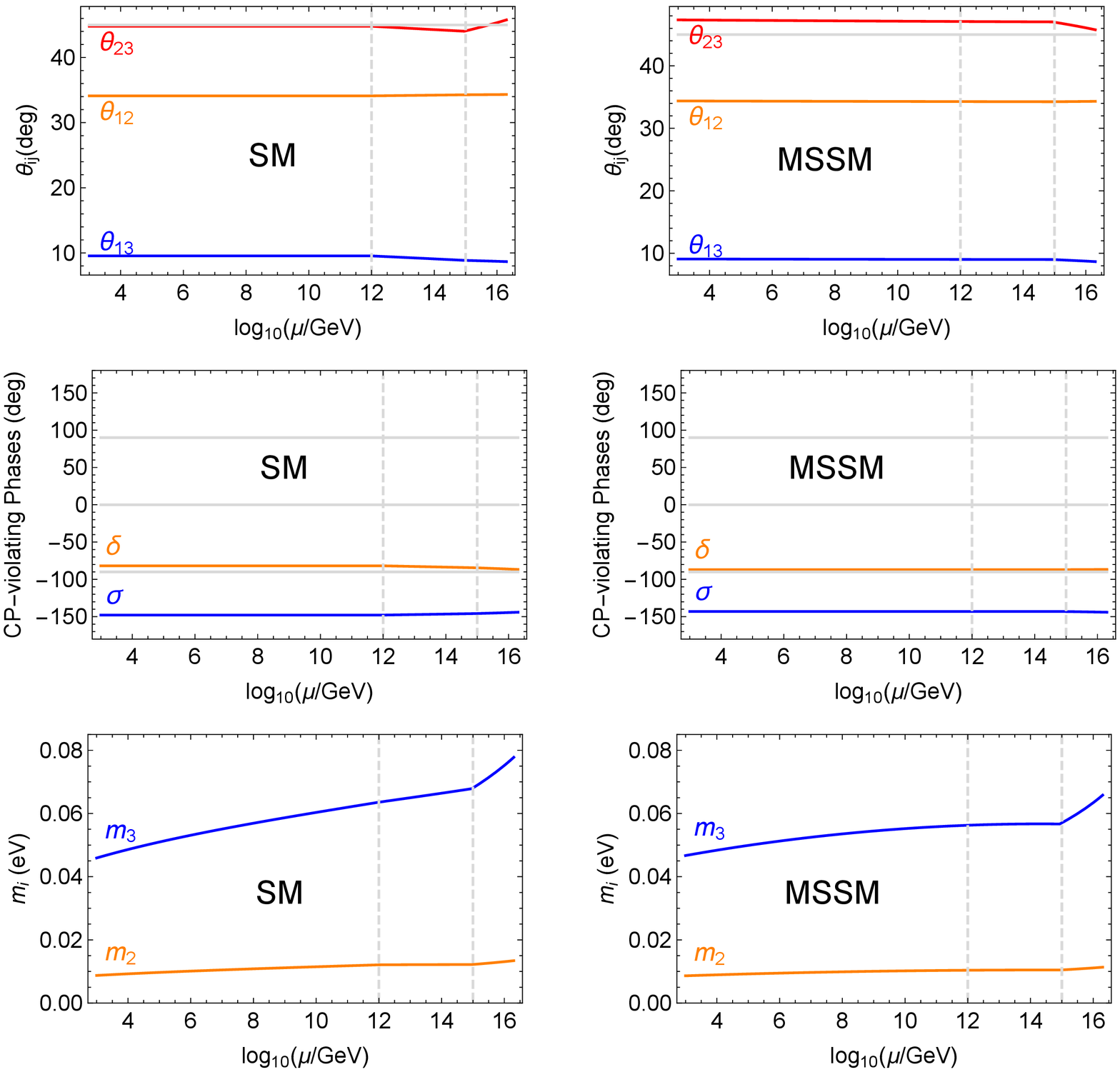}
\end{center}
\vspace{-0.3cm}
\caption{The evolution of three mixing angles $\{\theta^{}_{12}, \theta^{}_{13}, \theta^{}_{23}\}$, two CP-violating phases $\{\delta, \sigma\}$ and neutrino masses $\{m^{}_2, m^{}_3\}$ within the SM (left) and MSSM with $\tan\beta = 30$ (right) for the form of $Y_\nu^{}$ in {\bf Case C} given in Eq.~(\ref{eq:Ynu0I}). The initial values for the most relevant parameters at the high-energy scale $\mu^{}_0$ include $g_1^{} = 0.579$, $g_2^{} = 0.521$, $g_3^{} = 0.527$, $\lambda = 0.5$ (only for the SM), $y^{}_\tau = 0.010$ and $y_t^{} = 0.483$.}
\label{fg:fig3}
\end{figure}

\begin{figure}[!t]
\begin{center}
\includegraphics[width=\textwidth]{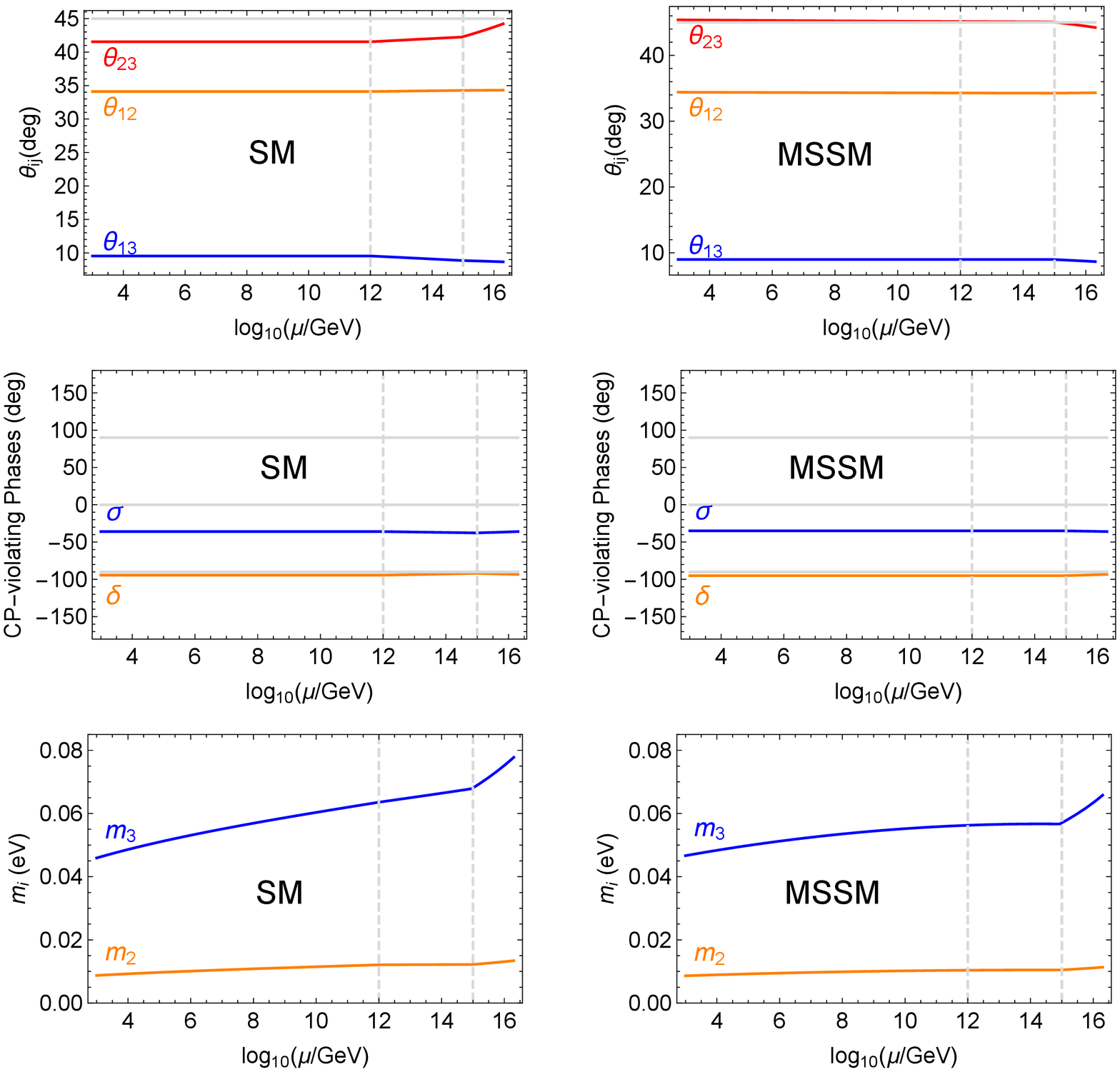}
\end{center}
\vspace{-0.3cm}
\caption{The evolution of three mixing angles $\{\theta^{}_{12}, \theta^{}_{13}, \theta^{}_{23}\}$, two CP-violating phases $\{\delta, \sigma\}$ and neutrino masses $\{m^{}_2, m^{}_3\}$ within the SM (left) and MSSM with $\tan\beta = 30$ (right) for the form of $Y_\nu^{}$ in {\bf Case D} given in Eq.~(\ref{eq:Ynu0I}). The initial values for the most relevant parameters at the high-energy scale $\mu^{}_0$ include $g_1^{} = 0.579$, $g_2^{} = 0.521$, $g_3^{} = 0.527$, $\lambda = 0.5$ (only for the SM), $y^{}_\tau = 0.010$ and $y_t^{} = 0.483$.}
\label{fg:fig4}
\end{figure}

Instead of an analytical approach, we adopt the exact numerical approach to solve the RG equations and show the final results in Figs.~\ref{fg:fig3} and \ref{fg:fig4} for cases C and D, respectively. The values at various energy scales are summarized in Tables~\ref{tb:tb4} and \ref{tb:tb5}. Note that the same scaling factor of 1.25 (1.15) has been applied to $Y_\nu^{}$ for SM (MSSM) so as to obtain better agreement with low-energy data on neutrino masses. Some comments on the numerical results are in order:
\begin{itemize}
\item Now we have more significant running effects on $\theta_{13}^{}$ and $\theta_{23}^{}$. For the previous ordering $M^{}_{\rm atm} \ll M^{}_{\rm sol}$, the running for $\theta^{}_{13}$ and $\theta^{}_{23}$ is about $0.5^\circ$ and $1.0^\circ$ for case A, respectively. The change of $\theta^{}_{23}$ for case B is even smaller, as indicated in Table~\ref{tb:tb3}. In the case of $M^{}_{\rm atm} \gg M^{}_{\rm sol}$, as shown in Table~\ref{tb:tb4}, both $\theta^{}_{13}$ and $\theta^{}_{23}$ get changed by about $1.0^\circ$ for case C. However, for case D, the results of $\theta^{}_{23}$ have been given in Table~\ref{tb:tb5}, and the decrease of $\theta^{}_{23}$ about $3^\circ$ is found for the SM, although the corrections in the MSSM are again small.

\item Regarding the running of $\theta^{}_{23}$ from $\Lambda^{}_{\rm GUT}$ to $M^{}_2$ in the SM, one can observe from Tables~\ref{tb:tb4} and \ref{tb:tb5} that the values of $\theta^{}_{23}$ decrease by about $2.0^\circ$, which is consistent with our expectation from Eq.~(\ref{eq:HnuI}). However, in the second stage from $M^{}_2$ to $M^{}_1$, $\theta^{}_{23}$ becomes increasing in case C, while it continues decreasing in case D. This opposite running behaviour may be ascribed to the competition among different contributions from both neutrino and charged-lepton sectors.
    
\item When running towards low energies, the ratio of $m_2^{}/m_3^{}$ becomes increasing while in the previous case it is decreasing. Moreover, the running of such a ratio is also more appreciable, and becomes in contradiction with the data. This can be attributed to a more significant running of $m_3^{}$. In principle, we can adjust both $m_a^{}$ and $m_b^{}$ such that neutrino masses are in good agreement with data, and even the tension of mixing angles with observations may also get reduced. For this purpose, a complete scan of model parameters should be carried out, which however is beyond the scope of the present work.
\end{itemize}

It is very interesting to notice that a deviation of $\theta^{}_{23}$ from the maximal mixing by $3^\circ$ can only be realised in the case of $M^{}_{\rm atm} \gg M^{}_{\rm sol}$ and the flavour structure of $Y^{}_\nu$ takes the form case D in Eq.~(\ref{eq:Ynu0I}). In the other cases, we are left with a nearly maximal mixing $\theta^{}_{23} = 45^\circ \pm 1^\circ$, including the radiative corrections.

\begin{table}
\footnotesize
\centering
\begin{tabular}{c | c  c  c  c | c  c  c  c | c}
\hline
\hline
& \multicolumn{4}{| c |}{SM} & \multicolumn{4}{ c|}{MSSM ($\tan\beta = 30$)} & \multirow{2}{*}{Best fit} \\
\cline{1-9}
& $\Lambda_{\mathrm{GUT}}^{}$ & $M_2^{}$ & $M_1^{}$ & $\Lambda_{\mathrm{EW}}^{}$ & $\Lambda_{\mathrm{GUT}}^{}$ & $M_2^{}$ & $M_1^{}$ & $\Lambda_{\mathrm{EW}}^{}$ & \\
\hline
$\theta_{13}^{} (\mathrm{deg})$ & 8.67 & 8.85 & 9.54 & 9.54 & 8.67 & 8.98 & 9.02 & 9.09 & $8.46^{+0.14}_{-0.15}$\\
$\theta_{12}^{} (\mathrm{deg})$ & 34.32 & 34.27 & 34.11 & 34.11 & 34.32 & 34.26 & 34.28 &  34.38 & $33.72^{+0.79}_{-0.76}$\\
$\theta_{23}^{} (\mathrm{deg})$ & 45.77 & 44.08 & 44.79 & 44.79 & 45.77  & 46.98 & 47.09 & 47.35  & $41.5^{+1.3}_{-1.1}$\\
\hline
$\delta (\mathrm{deg})$ & $- 86.7$ & $-84.5$ & $-81.8$ & $-81.8$ & $-86.7$ & $-86.9$ & $-86.9$ &  $-86.9$ & $-71^{+38}_{-51}$\\
$\sigma (\mathrm{deg})$ & $-144.0$ & $-145.8$ & $-147.8$ & $-143.2$ & $-144.0$ & $-143.1$ & $-143.1$ &  $-143.1$ & --\\
\hline
$m_2^{} (\mathrm{meV})$ & 13.4 & 12.2 & 12.1 & 8.75 & 11.4 & 10.5 & 10.4 &  8.63 & $8.65^{+0.11}_{-0.09}$\\
$m_3^{} (\mathrm{meV})$ & 77.8 & 68.1 & 63.6 & 45.9 & 65.8 & 57.0 & 56.3 &  46.7 & $50.26^{+0.39}_{-0.37}$\\
$m_2^{}/m_3^{}$ & 0.172 & 0.179 & 0.190 & 0.190 & 0.173 & 0.184 & 0.185 & 0.185 & $0.172^{+0.003}_{-0.003}$ \\
\hline
\hline
\end{tabular}
\caption{Three mixing angles $\{\theta^{}_{12}, \theta^{}_{13}, \theta^{}_{23}\}$, two CP-violating phases $\{\delta, \sigma\}$ and non-zero light neutrino masses $\{m^{}_2, m^{}_3\}$ at various energy scales according to two scenarios in {\bf Case C} given in Fig.~\ref{fg:fig3}. The best-fit results from Ref.~\cite{global1} are shown in the last column.}
\label{tb:tb4}
\end{table}

\begin{table}[!t]
\small
\centering
\begin{tabular}{c | c  c  c  c | c  c  c  c | c}
\hline
\hline
& \multicolumn{4}{| c |}{SM} & \multicolumn{4}{ c|}{MSSM ($\tan\beta = 30$)} & \multirow{2}{*}{Best fit} \\
\cline{1-9}
& $\Lambda_{\mathrm{GUT}}^{}$ & $M_2^{}$ & $M_1^{}$ & $\Lambda_{\mathrm{EW}}^{}$ & $\Lambda_{\mathrm{GUT}}^{}$ & $M_2^{}$ & $M_1^{}$ & $\Lambda_{\mathrm{EW}}^{}$ & \\
\hline
$\theta_{13}^{} (\mathrm{deg})$ & 8.67 & 8.85 & 9.54 & 9.54 & 8.67 & 8.98 & 8.99 & 8.99 & $8.46^{+0.14}_{-0.15}$\\
$\theta_{12}^{} (\mathrm{deg})$ & 34.32 & 34.27 & 34.11 & 34.11 & 34.32 & 34.26 & 34.29 &  34.40 & $33.72^{+0.79}_{-0.76}$\\
$\theta_{23}^{} (\mathrm{deg})$ & 44.22 & 42.30 & 41.54 & 41.54 & 44.22  & 45.04 & 45.14 & 45.39  & $41.5^{+1.3}_{-1.1}$\\
\hline
$\delta (\mathrm{deg})$ & $-93.3$ & $-92.0$ & $-94.4$ & $-94.4$ & $-93.3$ & $-95.0$ &  $-95.0$ & $-95.1$ & $-71^{+38}_{-51}$\\
$\sigma (\mathrm{deg})$ & $-36.0$ & $-37.7$ & $-36.0$ & $-36.0$ & $-36.0$ & $-35.0$ & $-35.0$ &  $-35.0$ & --\\
\hline
$m_2^{} (\mathrm{meV})$ & 13.4 & 12.2 & 12.1 & 8.75 & 11.4 & 10.5 & 10.4 &  8.64 & $8.65^{+0.11}_{-0.09}$\\
$m_3^{} (\mathrm{meV})$ & 77.8 & 68.1 & 63.6 & 45.9 & 65.8 & 57.0 & 56.3 &  46.7 & $50.26^{+0.39}_{-0.37}$\\
$m_2^{}/m_3^{}$ & 0.172 & 0.179 & 0.190 & 0.190 & 0.173 & 0.184 & 0.185 & 0.185 & $0.172^{+0.003}_{-0.003}$ \\
\hline
\hline
\end{tabular}
\caption{Three mixing angles $\{\theta^{}_{12}, \theta^{}_{13}, \theta^{}_{23}\}$, two CP-violating phases $\{\delta, \sigma\}$ and non-zero light neutrino masses $\{m^{}_2, m^{}_3\}$ at various energy scales according to two scenarios in {\bf Case D} given in Fig.~\ref{fg:fig4}. The best-fit results from Ref.~\cite{global1} are shown in the last column.}
\label{tb:tb5}
\end{table}

\subsection{Varying $M_{\rm atm}^{}$ and $M_{\rm sol}^{}$}

\begin{figure}
\includegraphics[scale=0.7]{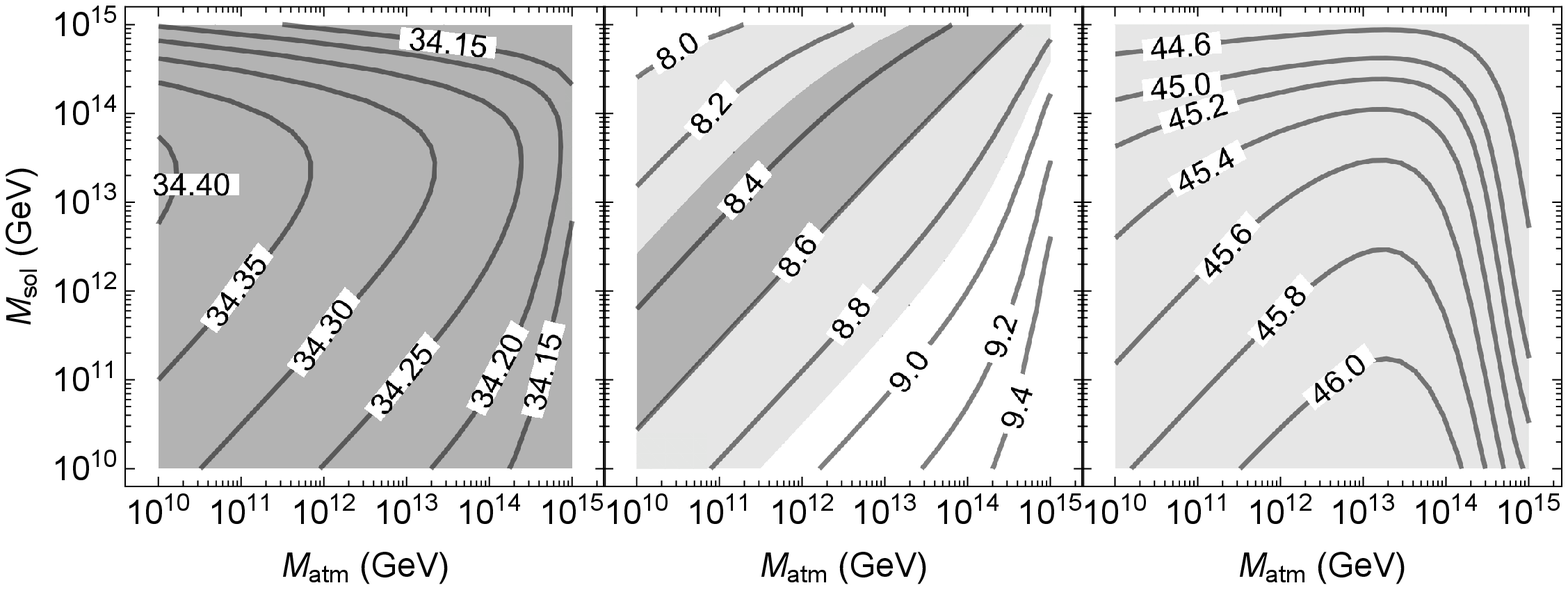}\\
\\
\includegraphics[scale=0.7]{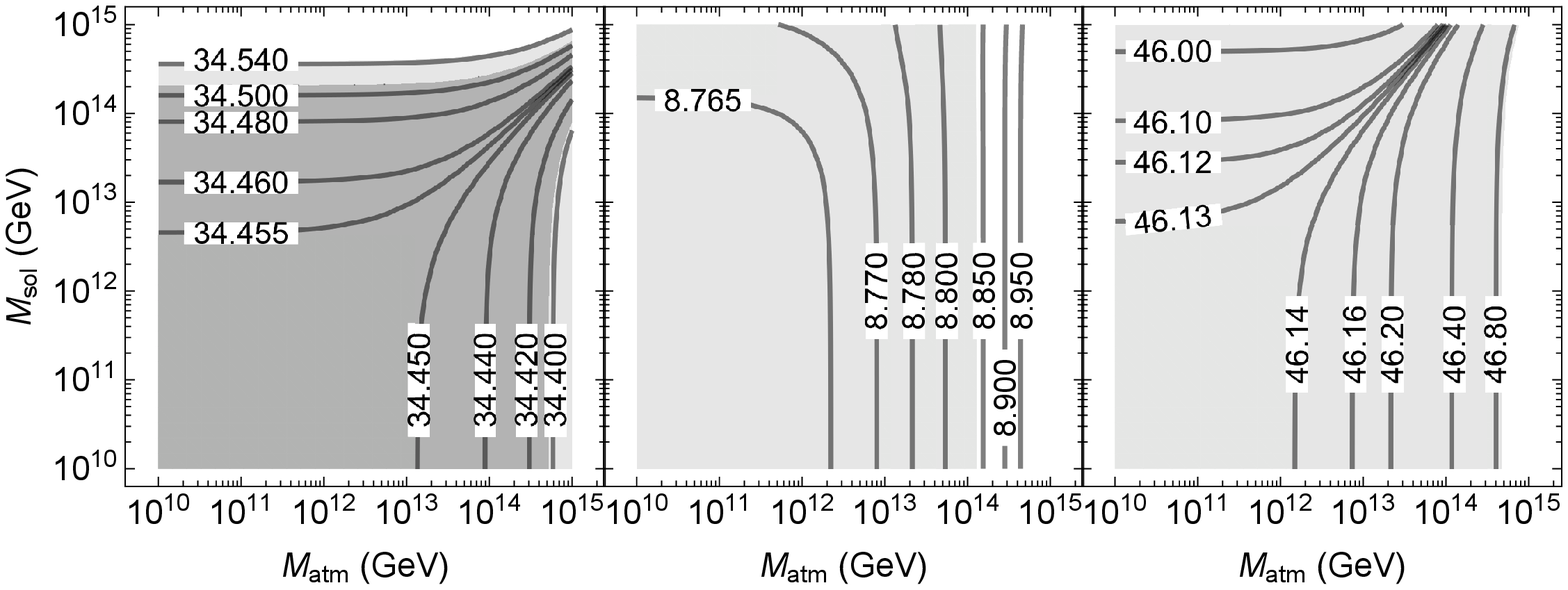}
\caption{Predicted mixing angles $\theta_{12}^{}$ (left), $\theta_{13}^{}$ (middle) and $\theta_{23}^{}$ (right) at the $\Lambda_{\mathrm{EW}}^{} = 10^3_{} ~\mathrm{GeV}$ in \textbf{Case A/C} within the cases of SM (top panel) and MSSM with $\tan\beta = 30$ (bottom panel), allowing both $M_{\mathrm{atm}}^{}$ and $M_{\mathrm{sol}}^{}$ to vary between  $10^{10}_{}~\mathrm{GeV}$ and  $10^{15}_{}~\mathrm{GeV}$. Boundary values of other model parameters, which yield $\theta_{13}^{} = 8.67^\circ_{}$, $\theta_{12}^{} = 34.32^\circ_{}$ and $\theta_{23}^{} = 45.77^\circ_{}$ at $\Lambda_{\mathrm{GUT}}^{}$, are chosen to be the same as those in Tables~\ref{tb:tb2} and \ref{tb:tb4}. Dark and light gray areas respectively correspond to $1\sigma$ and $3\sigma$ allowed regions, according to the global-fit results in Ref.~\cite{global1}.}
\label{fg:CSD_SM_MSSM}
\end{figure}

Finally, let us further expand our work to the scenario where both $M_{\rm atm}^{}$ and $M_{\rm sol}^{}$ are allowed to vary within certain ranges. We address this issue by evolving RG equations numerically, and choose the same boundary values of model parameters as those in Tables \ref{tb:tb2}-\ref{tb:tb5}, while varying both $M_{\mathrm{atm}}^{}$ and $M_{\mathrm{sol}}^{}$ between $10^{10}_{}~\mathrm{GeV}$ and  $10^{15}_{}~\mathrm{GeV}$. The obtained results for the form of $Y_\nu^{}$ as in \textbf{Case A} of Eq.~(\ref{eq:Ynu0}) and \textbf{Case C} of Eq.~(\ref{eq:Ynu0I}) are presented in Fig.~\ref{fg:CSD_SM_MSSM}, where both the cases of SM and MSSM with $\tan\beta = 30$ are considered. To compare with the current experimental data, we also include the $1\sigma$ and $3\sigma$ allowed regions according to the global-fit results in Ref.~\cite{global1}. Several observations are then made:
\begin{itemize}
\item In the entirely considered ranges of $M_{\mathrm{atm}}^{}$ and $M_{\mathrm{sol}}^{}$, the running effects for the three mixing angles are all rather small, at most one degree for $\theta_{13}^{}$ and $\theta_{23}^{}$. 

\item In comparison with the global-fit results, we see that having $\theta_{12}^{}$ to be compatible with the data, even at the level of $1\sigma$, is easy to achieve. However, for $\theta_{13}^{}$ and $\theta_{23}^{}$, although a $3\sigma$ level of agreement is also not difficult, reaching a compatibility at the $1\sigma$ level becomes impossible in $\theta_{23}^{}$, and for $\theta_{13}^{}$ it is only in the case of SM that there exists some parameter space of $M_{\mathrm{atm}}^{}$ and $M_{\mathrm{sol}}^{}$. 
\end{itemize}
It should be pointed out that in Fig.~\ref{fg:CSD_SM_MSSM} we also consider the case where $M_{\mathrm{atm}}^{}$ and $M_{\mathrm{sol}}^{}$ are almost degenerate, while the previously derived analytical results are only applicable to the hierarchical cases. 

\begin{figure}
\includegraphics[scale=0.7]{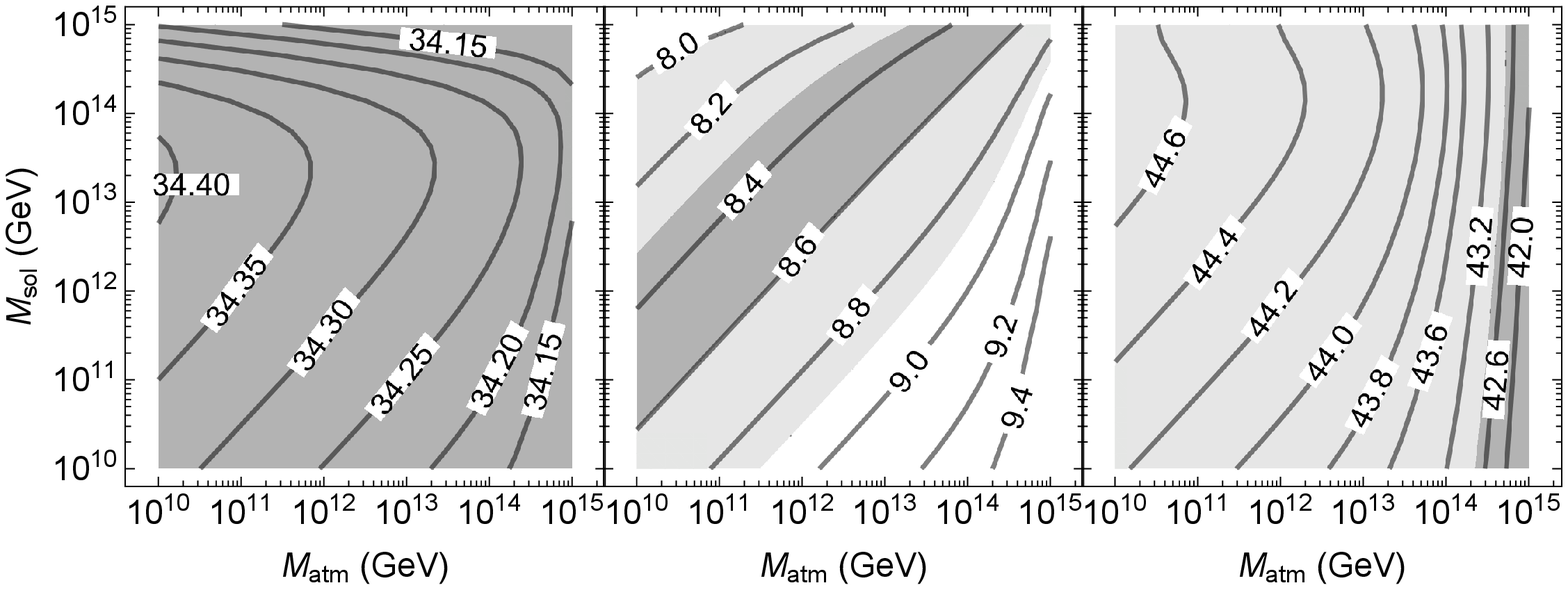}\\
\\
\includegraphics[scale=0.7]{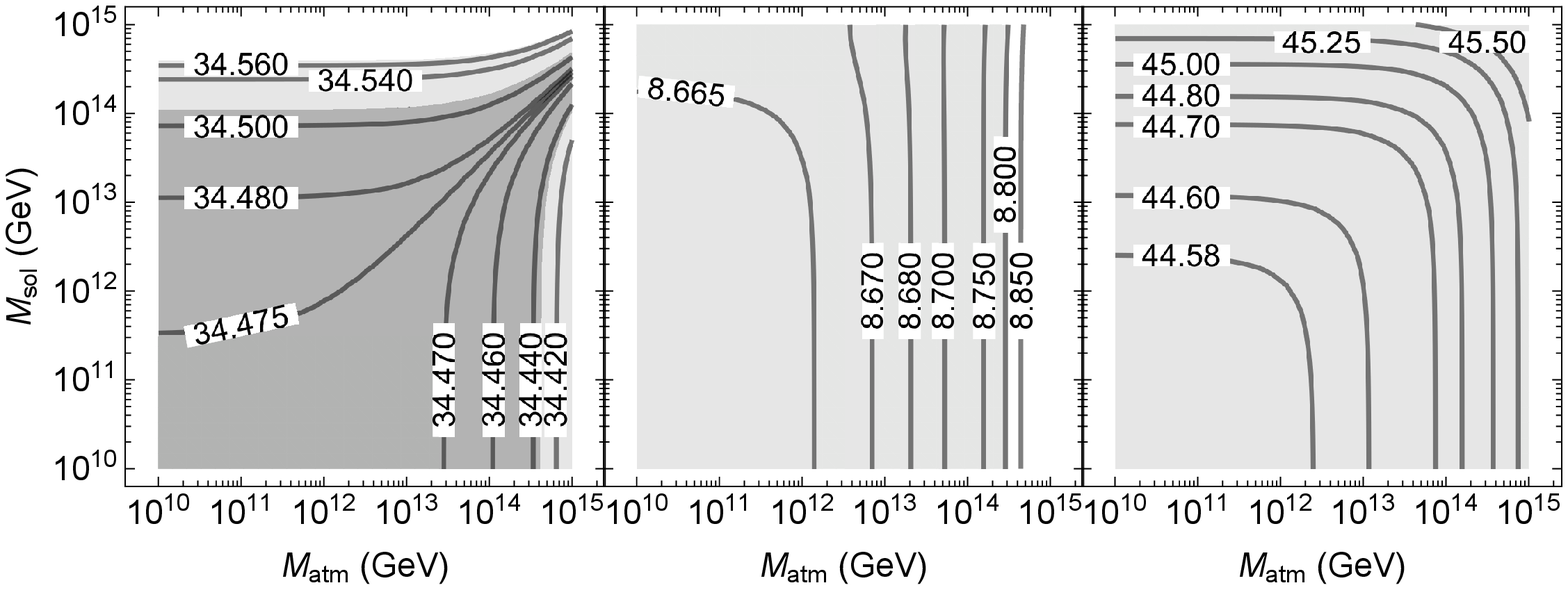}
\caption{Predicted mixing angles $\theta_{12}^{}$ (left), $\theta_{13}^{}$ (middle) and $\theta_{23}^{}$ (right) at the $\Lambda_{\mathrm{EW}}^{} = 10^3_{} ~\mathrm{GeV}$ in \textbf{Case B/D} within the cases of SM (top panel) and MSSM with $\tan\beta = 30$ (bottom panel), allowing both $M_{\mathrm{atm}}^{}$ and $M_{\mathrm{sol}}^{}$ to vary between  $10^{10}_{}~\mathrm{GeV}$ and  $10^{15}_{}~\mathrm{GeV}$. Boundary values of other model parameters, which yield $\theta_{13}^{} = 8.67^\circ_{}$, $\theta_{12}^{} = 34.32^\circ_{}$ and $\theta_{23}^{} = 44.22^\circ_{}$ at $\Lambda_{\mathrm{GUT}}^{}$, are chosen to be the same as those in Tables~\ref{tb:tb3} and \ref{tb:tb5}. Dark and light gray areas respectively correspond to $1\sigma$ and $3\sigma$ allowed regions, according to the global-fit results in Ref.~\cite{global1}.}
\label{fg:alter_CSD_SM_MSSM}
\end{figure}

We then turn to the other form of $Y_\nu^{}$, namely, \textbf{Case B} in Eq.~(\ref{eq:Ynu0}) and \textbf{Case D} of Eq.~(\ref{eq:Ynu0I}). In Fig.~\ref{fg:alter_CSD_SM_MSSM} we show the numerical results obtained in the same way as the above. As one can see, the running of three mixing angle is again rather insignificant, except for $\theta_{23}^{}$ in the case of SM, for which the decrease of $\theta_{23}^{}$ around $3^\circ$ can appear when $M_{\mathrm{atm}}^{} \sim 10^{15}_{}~\mathrm{GeV}$ as discussed in the previous section. Therefore, we have extended our previous conclusion, i.e., in the presence of radiative corrections a close to maximal atmospheric mixing of $\theta_{23}^{} = 45^\circ \pm 1^\circ$ can be achieved in most of cases,  to the scenario that both $M_{\mathrm{atm}}^{}$ and $M_{\mathrm{sol}}^{}$ are varied in a wide range. Such a robust prediction on $\theta_{23}^{}$ calls for scrutiny under future neutrino experimental results.

\section{Summary}

Seesaw models are able to explain simultaneously both tiny neutrino masses and the cosmological matter-antimatter asymmetry,
but generally involve a large number of parameters.
By contrast, the LS model involves two right-handed neutrinos and a very constrained Dirac mass matrix, involving one texture zero and two independent Dirac masses, leading to a highly predictive scheme in which all neutrino masses and the entire PMNS matrix is
successfully predicted in terms of just two real parameters. To be precise, we have considered two simple structures of the Dirac neutrino Yukawa coupling matrix $Y^{}_\nu$, denoted as cases A and B, with $M^{}_{\rm atm}\ll M^{}_{\rm sol}$
each of which contains only three real parameters $a$, $b$, and $\eta$, which may be fixed by symmetry arguments to be a cube root of unity, leading to testable predictions for low-energy neutrino experiments.
We also considered two related cases C and D corresponding to $M^{}_{\rm sol}\ll M^{}_{\rm atm}$.

Each case predicts a
normal neutrino mass hierarchy with $\{m^{}_1, m^{}_2, m^{}_3\} = \{0, 8.6, 50\}$ meV,
where the effective neutrino mass $m^{}_{\beta \beta} = 2.7~{\rm meV}$ for neutrinoless double-beta decays is so small that observation of such decays is impossible in the foreseeable future. The LS model also predicts an almost maximal CP-violating phase
$\delta = -87^\circ$ (cases A,C) or $-93^\circ$ (cases B,D) which will be verified or ruled out in the future oscillation experiments.
The LS model also predicts close to maximal atmospheric mixing at the high scale,
$\theta_{23}\approx 46^\circ $ (cases A,C), or $\theta_{23}\approx 44^\circ $ (cases B,D), where both
predictions are challenged by the latest NOvA results in the $\nu_{\mu}$ disappearance channel
which indicates that $\theta_{23}=45^\circ$
is excluded at the 2.5 $\sigma$ CL, although T2K measurements in the same channel continue to prefer maximal mixing.

In this paper, motivated by the simplicity and predictivity of the LS, we have
calculated the RG corrections to the LS predictions, for both cases A and B, with and without supersymmetry,
including also the threshold effects induced by the decoupling of heavy Majorana neutrinos both analytically and numerically.
We also performed a numerical RG analysis for cases C and D.
In particular
we have investigated the RG running of three neutrino mixing angles, taking account of the threshold effects induced by the decoupling of heavy Majorana neutrinos, including both possible mass orderings of right-handed neutrinos.
Although the running effects are rather small both in the SM and in the MSSM with $\tan \beta = 30$, we have carried out an analytical treatment of the RG running between two seesaw thresholds for cases A,B. We emphasise that the full numerical calculation was performed to verify our analytical and approximate results.
We find that the predictions for neutrino mixing angles
and mass ratios are rather stable under RG corrections. For example we find that
the LS model with RG corrections always predicts close to maximal atmospheric mixing $\theta_{23}=45^\circ \pm 1^\circ$,
for most considered cases, which remains in tension with the latest NOvA results. The one exception is case D for the SM, where
$\theta_{23}= 41.5^\circ$ after RG corrections.

Finally we mention that the techniques used here may be applied to other seesaw models with a strong normal mass hierarchy.
We hope that such future studies would be helpful in revealing how the RG running modifies theoretical predictions for neutrino mixing parameters, for other related neutrino mass models with flavour symmetries.
In particular, the results here are expected to be indicative of a large class of seesaw models with a strong mass hierarchy that
predict close to maximal atmospheric mixing, so we conclude that RG corrections are not generally sufficient to rescue such models
if maximal atmospheric mixing becomes excluded.

\section*{Acknowledgements}

The authors would like to thank Zhi-zhong Xing for useful discussions and partial involvement at the early stage of this work, which was supported in part by the National Recruitment Program for Young Professionals and by the CAS Center for Excellence in Particle Physics (CCEPP).
The authors are also grateful to the Mainz Institute for Theoretical
Physics (MITP) where this work was started and for its hospitality and its partial support.
SFK acknowledges support from the STFC Consolidated grant ST/L000296/1 and the
European Union Horizon 2020 research and innovation programme under the Marie
Sklodowska-Curie grant agreements InvisiblesPlus RISE No. 690575 and
Elusives ITN No. 674896.

\newpage

\end{document}